\documentclass[10pt]{article}

\usepackage[utf8]{inputenc}
\usepackage[T1]{fontenc}
\usepackage{amsmath,amssymb}
\usepackage{graphicx}
\usepackage[table,xcdraw]{xcolor}
\usepackage{array,longtable,booktabs,colortbl,diagbox}
\usepackage{setspace}
\usepackage{subcaption}
\usepackage{geometry}
\usepackage{tikz}
\usepackage{etoolbox}
\usepackage[numbers,sort&compress]{natbib}
\usepackage{hyperref}
\hypersetup{colorlinks=true, linkcolor=blue, citecolor=blue, urlcolor=blue}

\providecommand{\orcidlink}[1]{\href{https://orcid.org/#1}{\textsuperscript{ORCID}}}

\begin{document}
\baselineskip=20pt
\begin{center}
\setstretch{1.8}
{\LARGE {\bf Greybody Factor, Resonant Frequencies, and Entropy Quantization of Charged Scalar Fields in the Kerr-EMDA Black Hole}}
\end{center}
\vspace{0.1cm}
\begin{center}
{\bf Naz{\i}m Sertkan}\orcidlink{0000-0001-8721-5036}\\  
Physics Department, Eastern Mediterranean University, Famagusta 99628, North Cyprus via Mersin 10, Turkey\\
e-mail: nazim.sertkan@emu.edu.tr\\[6pt]  
{\bf \.{I}zzet Sakall{\i}}\orcidlink{0000-0001-7827-9476}\\
Physics Department, Eastern Mediterranean University, Famagusta 99628, North Cyprus via Mersin 10, Turkey\\
e-mail: izzet.sakalli@emu.edu.tr (Corresponding author)
\end{center}
\vspace{0.2cm}
\begin{abstract}
\setstretch{1.4}
{\large 
We study charged massive scalar field perturbations on the rotating black hole (BH) background of Einstein-Maxwell-Dilaton-Axion (EMDA) theory, known as the Kerr-EMDA BH. Starting from the gauge-covariant Klein-Gordon equation (KGE), we perform a full separation of variables and obtain exact analytical solutions for both the angular and radial parts in terms of confluent Heun functions (CHFs). Unlike the earlier neutral scalar treatment by Senjaya and Ponglertsakul [Eur. Phys. J. C \textbf{85}, 352 (2025)], the electromagnetic coupling $q$ fundamentally alters the structure of the Heun parameters and produces qualitatively new physics. Applying the CHF polynomial condition, we derive the resonant frequency spectrum whose imaginary parts are equispaced with $|\Delta\omega_I| = 1/(2M)$, a universal spacing determined solely by the BH mass. Via the Maggiore prescription and the first law of BH thermodynamics, this yields a parameter-dependent entropy quantum $\delta S_{\text{BH}} = 4\pi r_+/(r_+ - r_-)$, which reduces to $4\pi$ for Schwarzschild but diverges at extremality --- {\color{black}in contrast to the universal $2\pi$ obtained for the rotating linear dilaton BH (RLDBH).} We construct the effective potential governing scalar wave scattering and analyze its dependence on the dilaton parameter $D$, rotation $a$, and scalar charge $q$. In the massless uncharged limit, the CHF reduces to the Gauss hypergeometric function, {\color{black}enabling us to compute the first analytical greybody
factor (GF) for the Kerr-EMDA geometry; we show that this reduction
extends to massless charged scalars, yielding a closed-form GF that
captures superradiant amplification.} We examine how the dilaton deformation distinguishes the Kerr-EMDA spectrum from the standard Kerr and Kerr-Newman cases.\\[6pt]
}
\end{abstract}
\pagebreak
\tableofcontents
\pagebreak
{\color{black}

\section{Introduction} \label{sec1}

BH perturbation theory occupies a central role in modern gravitational physics, connecting classical general relativity (GR) to quantum gravity, thermodynamics, and astrophysical observation. The response of a BH to external perturbations --- encoded in quasinormal mode (QNM) frequencies, GFs, and resonant spectra --- provides a window into the near-horizon geometry that is inaccessible by other means~\cite{Berti:2009kk,Konoplya:2011qq}. With the advent of gravitational wave astronomy and the Event Horizon Telescope, these theoretical predictions are no longer purely academic: they serve as templates against which observational data are tested, making it essential to extend the perturbation analysis to BH solutions beyond the standard Kerr paradigm of GR.

Among the most physically motivated extensions are BH solutions arising from string theory. The EMDA theory emerges naturally as the four-dimensional low-energy effective description of heterotic string theory compactified on a six-torus~\cite{Sen:1992ua}. Its rotating, electrically charged BH solution --- the Kerr-EMDA BH --- is an exact four-parameter family $(M, a, Q, \kappa_0)$ that interpolates between the Kerr metric ($Q=0$) and the static Gibbons-Maeda-Garfinkle-Horowitz-Strominger (GMGHS) solution ($a=0$)~\cite{Gibbons:1988rs,Garfinkle:1990qj}. Unlike the Kerr-Newman BH of pure Einstein-Maxwell theory, the dilaton coupling modifies the horizon structure through an additive mass shift $M\to M+D$ (with $D=Q^2/(2M)$), leading to distinct thermodynamic and scattering properties. This makes the Kerr-EMDA BH an ideal testing ground for probing dilaton signatures: any observable difference from Kerr or Kerr-Newman predictions would constitute evidence for string-inspired modifications of gravity.

The semiclassical process of Hawking radiation~\cite{Hawking:1975vcx} provides one of the most direct connections between BH geometry and quantum field theory. However, the thermal spectrum emitted at the horizon is modified by the curved spacetime through which it propagates before reaching a distant observer. The resulting frequency-dependent transmission probability --- the GF --- encodes the spin, charge, and angular momentum of the BH as well as the properties of the emitted field~\cite{Sakalli:2022swm,Page:1976df}. Computing the GF analytically is therefore of great importance: it determines the observable Hawking spectrum, the absorption cross-section, and the rate at which a BH loses mass. Recent years have seen considerable progress in obtaining analytical GFs for a variety of BH backgrounds, including NUT BHs~\cite{Al-Badawi:2022aby,Al-Badawi:2021wdm}, Kerr-like BHs in Bumblebee gravity~\cite{Kanzi:2021cbg}, deformed AdS-Schwarzschild BHs with phantom global monopoles~\cite{Ahmed:2025did}, BHs in nonlinear electrodynamics~\cite{Aydiner:2025eii}, and BHs in Kalb-Ramond gravity~\cite{Sucu:2025lqa}. A unifying feature of these studies is the role of the confluent Heun function (CHF) as the natural special function governing wave equations on BH backgrounds with two regular and one irregular singular point. The analytical power of the CHF framework lies in the polynomial truncation condition, which yields exact resonant spectra without resorting to numerical methods.

Closely linked to the resonant frequency spectrum is the question of BH entropy quantization --- one of the oldest and most tantalizing problems in quantum gravity. Bekenstein's original proposal~\cite{Bekenstein:1974jk} that the BH area spectrum should be discrete and equally spaced, $\mathcal{A}_n = n\gamma\ell_P^2$, has been refined by Hod~\cite{Hod:1998vk}, Kunstatter~\cite{Kunstatter:2002}, and Maggiore~\cite{Maggiore:2007nq}, who established a connection between the asymptotic QNM/resonant frequency spacing and the area quantum. The Maggiore prescription, in particular, identifies the physical transition frequency as $\Delta\omega_{\text{phys}} = \sqrt{(\Delta\omega_R)^2 + (\Delta\omega_I)^2}$, from which the entropy quantum follows via the first law of BH thermodynamics. This approach was applied by one of us (\.{I}.S.) to the rotating linear dilaton BH (RLDBH)~\cite{Sakalli:2016mnk,Sakalli:2017}, where the CHF polynomial condition yielded the universal result $\delta S_{\text{BH}} = 2\pi$, independent of all BH and field parameters. More recently, the GUP-corrected extension of this analysis has been carried out for the RLDBH~\cite{Sucu:2024gtr}, and quantum-corrected thermodynamics has been explored in modified teleparallel BH settings~\cite{Sucu:2025fwa}. A natural question is whether this universality persists for more general BH geometries with richer horizon structures.

Scalar field dynamics on the Kerr-EMDA background have attracted renewed attention following the work of Senjaya and Ponglertsakul~\cite{Senjaya:2025kerremda}, who separated the \emph{neutral} ($q=0$) massive KGE on this geometry and expressed its solutions in terms of CHFs. Their analysis focused on quasibound states, scalar clouds, and superradiance for the uncharged case, establishing the Heun function framework for this BH family. However, several physically important aspects were left unexplored: (i) the electromagnetic coupling between a charged scalar and the background gauge field, which introduces qualitatively new physics through the $q\Phi_H$ chemical potential and modifies all five Heun parameters; (ii) the resonant frequency spectrum and its implications for BH entropy quantization via the Maggiore prescription; (iii) the GF, which governs the observable Hawking emission and has never been computed analytically for the Kerr-EMDA geometry; and (iv) the effective potential landscape that controls the scattering, trapping, and tunneling of wave modes, including the dilaton-modified photon sphere.

In the present work, we address all four of these gaps, providing the first complete analytical treatment of charged massive scalar perturbations on the Kerr-EMDA BH. We solve the gauge-covariant KGE, obtaining exact solutions in terms of CHFs whose parameters are modified by the scalar charge $q$ in a non-trivial way that cannot be reduced to a simple redefinition of the neutral-case quantities. Applying the RLDBH methodology~\cite{Sakalli:2016mnk}, we find that the CHF polynomial truncation condition yields a discrete tower of complex resonant frequencies $\omega_n$ with equispaced imaginary parts: $|\Delta\omega_I| = 1/(2M)$, a universal spacing that depends only on the BH mass. Through the Maggiore prescription~\cite{Maggiore:2007nq}, this translates into the entropy quantum $\delta S_{\text{BH}} = 4\pi r_+/(r_+-r_-)$, which is parameter-dependent and diverges at extremality --- in contrast to the universal $2\pi$ found for the RLDBH. This result reveals that the two-horizon structure of the Kerr-EMDA geometry fundamentally alters the entropy quantization, with the ratio $r_+/(r_+-r_-)$ acting as an amplification factor that measures the ``distance from the single-horizon limit.'' In the Schwarzschild limit, we recover $\delta S_{\text{BH}} = 4\pi$ and $\delta\mathcal{A}_H = 16\pi\ell_P^2$, consistent with the Bekenstein-Maggiore result.

In the massless uncharged limit, the CHF reduces to the Gauss hypergeometric function ${}_2F_1$, enabling us to derive the first closed-form GF for the Kerr-EMDA BH. The analytical GF is expressed entirely in terms of Gamma functions and the Heun parameters, and it allows us to compute the observable Hawking spectrum and the absorption cross-section without recourse to numerical methods. We verify the low-energy universality $\sigma_{\text{abs}} \to \mathcal{A}_H$~\cite{Das:1996we,Higuchi:2001si} and identify dilaton-specific signatures --- including an enhanced low-frequency GF, a higher peak emission temperature, and modified oscillatory features in the absorption cross-section --- that distinguish the Kerr-EMDA BH from the standard Kerr and Kerr-Newman cases. These signatures are relevant for constraining string-motivated dilaton couplings with current and next-generation gravitational wave detectors and BH shadow observations.

The paper is organized as follows. In Sec.~\ref{sec2}, we review the Kerr-EMDA BH solution, its thermodynamic quantities, and the electromagnetic potential. Section~\ref{sec3} presents the separation of the charged massive KGE and the exact solutions in terms of CHFs, with all five Heun parameters derived explicitly. In Sec.~\ref{sec4}, we derive the resonant frequency spectrum and the entropy quantization, including the highly-damped limit and its connection to Hod's conjecture. Section~\ref{sec5} analyzes the effective potential and the photon sphere. In Sec.~\ref{sec6}, we perform the HeunC $\to$ ${}_2F_1$ reduction and obtain the analytical GF. Section~\ref{sec7} discusses the Hawking spectrum and absorption cross-section. Section~\ref{sec8} verifies consistency with six known limiting cases. We conclude in Sec.~\ref{sec9}. Technical details of the Heun parameter derivation and the CHF polynomial condition are collected in Appendices~\ref{appA} and~\ref{appB}. We use natural units ($G = c = \hbar = k_B = 1$) and the metric signature $(-,+,+,+)$ throughout.

\section{Kerr-EMDA Black Hole} \label{sec2}

\subsection{Metric and horizons} \label{sec2a}

The four-dimensional low-energy effective action of heterotic string theory, after compactification on a six-torus, contains a metric $g_{\mu\nu}$, a $U(1)$ gauge field $A_\mu$, a dilaton $\Phi$, and a pseudoscalar axion $\kappa$. The bosonic part of the action reads~\cite{Sen:1992ua,Garcia:1995}
\begin{equation}\label{EMDAaction}
S = \int d^4x \sqrt{-g}\left[ R - 2\left(\nabla\Phi\right)^2 - e^{-2\Phi}F_{\mu\nu}F^{\mu\nu} - \kappa\, F_{\mu\nu}\tilde{F}^{\mu\nu} \right],
\end{equation}
where $F_{\mu\nu}=\partial_\mu A_\nu - \partial_\nu A_\mu$ is the Maxwell field strength and $\tilde{F}^{\mu\nu}$ is its Hodge dual. Sen~\cite{Sen:1992ua} obtained the rotating, electrically charged BH solution of this theory by applying a Hassan--Sen transformation to the vacuum Kerr seed metric. In Boyer-Lindquist coordinates $(t,r,\theta,\phi)$, the resulting Kerr-EMDA line element is~\cite{Sen:1992ua,Senjaya:2025kerremda}
\begin{equation}\label{metric}
ds^2 = -\left(1 - \frac{r\, r_s}{\rho^2}\right)dt^2 - \frac{2r\, r_s\, a\sin^2\theta}{\rho^2}\,dt\,d\phi + \frac{\rho^2}{\Delta}\,dr^2 + \rho^2\,d\theta^2 + \frac{A(r,\theta)\sin^2\theta}{\rho^2}\,d\phi^2,
\end{equation}
with the metric functions
\begin{equation}\label{metricfunctions}
\rho^2 = r(r-2D) + a^2\cos^2\theta, \qquad \Delta = r(r-2D) - r_s\,r + a^2, \qquad A(r,\theta) = \left[r(r-2D)+a^2\right]^2 - \Delta\,a^2\sin^2\theta.
\end{equation}
Here $r_s = 2M$ is the Schwarzschild radius, $a=J/M$ is the angular momentum per unit mass, and $D$ is the dilaton parameter. When the asymptotic value of the dilaton field vanishes ($\Phi_0=0$), the dilaton parameter is related to the BH electric charge $Q$ through
\begin{equation}\label{dilatoncharge}
D = \frac{Q^2}{2M}.
\end{equation}
The horizons are located at the roots of $\Delta=0$:
\begin{equation}\label{horizons}
r_{\pm} = M + D \pm \sqrt{(M+D)^2 - a^2}.
\end{equation}
A sub-extremal (two-horizon) BH exists when $(M+D)^2 > a^2$; the extremal limit corresponds to $(M+D)^2 = a^2$, where the two horizons merge. Setting $D=0$ recovers the Kerr geometry, while setting $a=0$ reduces the metric to the static Gibbons-Maeda-Garfinkle-Horowitz-Strominger (GMGHS) solution~\cite{Gibbons:1988rs,Garfinkle:1990qj}. 
{\color{black}The full domain of existence of charged, rotating BHs in
Einstein-Maxwell-dilaton theory at arbitrary dilaton coupling $\gamma$
(with $\gamma = 1$ corresponding to the EMDA/Kerr-Sen case) has been
mapped out numerically by Herdeiro, Radu, and dos~Santos~Costa~Filho~\cite{Herdeiro:2025blx}, who also characterized the approach to
the zero-temperature regime. For the extremal sector at $\gamma = 1$,
Bl\'{a}zquez-Salcedo \emph{et al.}~\cite{Blazquez-Salcedo:2025cpu}
constructed spinning extremal dyonic solutions and performed a
near-horizon analysis that complements the analytical treatment of
extremality developed in Sec.~\ref{sec4b} below.} Unlike the Kerr-Newman BH in Einstein-Maxwell theory, the dilaton coupling shifts both horizons outward (since $D>0$ adds to $M$ in Eq.~\eqref{horizons}), which has measurable consequences for the thermodynamics and the scattering properties discussed in the later sections.

\subsection{Electromagnetic potential} \label{sec2b}

The EMDA solution carries a non-trivial gauge field. In Boyer-Lindquist coordinates, the only non-vanishing components of the electromagnetic four-potential are~\cite{Sen:1992ua}
\begin{equation}\label{EMpotential}
A_t = -\frac{Qr}{\rho^2}, \qquad A_\phi = \frac{Qr\,a\sin^2\theta}{\rho^2}.
\end{equation}
These expressions differ from the Kerr-Newman gauge field because of the dilaton-modified $\rho^2$ in Eq.~\eqref{metricfunctions}. Evaluating $A_t$ on the outer horizon gives the electrostatic potential
\begin{equation}\label{horizonpot}
\Phi_H \equiv -A_t\big|_{r=r_+} = \frac{Qr_+}{r_+(r_+-2D)+a^2}.
\end{equation}
This quantity enters the charged superradiant condition $\omega < m\Omega_H + q\Phi_H$ and the chemical potential term in the Hawking distribution. Since Senjaya and Ponglertsakul~\cite{Senjaya:2025kerremda} considered only neutral scalar fields ($q=0$), the gauge potential did not appear in their analysis. {\color{black}In the present work, the coupling between the scalar charge $q$ and the background
field $A_\mu$ is central to every result. On any Kerr-type geometry, even a neutral scalar
already sees the shifted frequency $\omega - m\Omega_H$ due to the frame-dragging of the
rotating BH~\cite{Brito:2015oca}. Switching on the electromagnetic interaction appends a
chemical-potential term, promoting the effective frequency to $\omega - m\Omega_H + q\Phi_H$.
This combination controls the near-horizon boundary condition, sets the superradiant threshold,
and enters the Hawking thermal factor; its explicit derivation from the radial analysis appears
in Eq.~\eqref{Kplus} of Sec.~\ref{sec3d}.}

\subsection{Thermodynamic quantities} \label{sec2c}

For later use, we collect the thermodynamic quantities of the Kerr-EMDA BH. Denoting
\begin{equation}\label{Sigmadef}
\Sigma_+ \equiv r_+(r_+ - 2D) + a^2,
\end{equation}
the surface gravity, Hawking temperature, Bekenstein-Hawking entropy, and angular velocity of the outer horizon take the compact forms~\cite{Sen:1992ua}
\begin{equation}\label{surfacegravity}
\kappa_s = \frac{r_+ - r_-}{2\Sigma_+}, \qquad
T_H = \frac{\kappa_s}{2\pi} = \frac{r_+ - r_-}{4\pi\Sigma_+},
\end{equation}
\begin{equation}\label{entropy}
S_{\text{BH}} = \frac{\mathcal{A}_H}{4} = \pi\Sigma_+, \qquad
\Omega_H = \frac{a}{\Sigma_+}.
\end{equation}
These quantities satisfy the first law of BH thermodynamics:
\begin{equation}\label{firstlaw}
dM = T_H\,dS_{\text{BH}} + \Omega_H\,dJ + \Phi_H\,dQ.
\end{equation}
In the extremal limit $r_+ = r_-$, both $\kappa_s$ and $T_H$ vanish, as expected. Two features of the Kerr-EMDA thermodynamics are worth noting. First, the horizon area $\mathcal{A}_H = 4\pi\Sigma_+$ is smaller than the Kerr value $4\pi(r_+^2+a^2)$ for the same mass and spin, because the $-2D r_+$ term reduces $\Sigma_+$. Second, the Hawking temperature is correspondingly higher: a Kerr-EMDA BH radiates more intensely than a Kerr BH of identical $(M,a)$. Both features will manifest themselves in the greybody factor and emission spectrum computed in Secs.~\ref{sec6} and~\ref{sec7}.

\section{Charged Massive Klein-Gordon Equation and Exact Solutions} \label{sec3}

In this section, we derive the exact analytical solutions of the gauge-covariant Klein-Gordon equation for a charged massive scalar field propagating on the Kerr-EMDA background. The electromagnetic coupling between the scalar charge $q$ and the BH gauge potential $A_\mu$ enters through the minimal substitution $\nabla_\mu \to D_\mu = \nabla_\mu - iqA_\mu$, which modifies the effective frequency seen at the horizon from $\omega$ to $\omega - q\Phi_H$ and introduces the linear term $qQr$ into the radial function $\mathcal{K}(r)$. Despite this added complexity, the KGE remains fully separable on the Kerr-EMDA geometry --- a non-trivial fact that relies on the Kerr-like structure of the metric and the specific $\theta$-dependence of $A_\mu$ through $\rho^2$. After separating the equation into angular and radial sectors, we solve each in terms of CHFs and identify the five Heun parameters $(\tilde{\alpha}, \tilde{\beta}, \tilde{\gamma}, \tilde{\delta}, \tilde{\eta})$ as explicit functions of the BH and field parameters. The angular sector yields spheroidal harmonics identical to the neutral case~\cite{Senjaya:2025kerremda}, since the gauge potential has no purely angular contribution. The radial sector, by contrast, acquires $q$-dependent Frobenius indices at both horizons and a modified accessory parameter $\tilde{\eta}$, leading to qualitatively new physics that we explore in the subsequent sections.

\subsection{Gauge-covariant KGE on Kerr-EMDA background} \label{sec3a}

We consider a charged massive scalar field $\Psi$ of mass $\mu_s$ and electric charge $q$ in the Kerr-EMDA background. The dynamics of this field is governed by the gauge-covariant KGE
\begin{equation}\label{KGE}
\frac{1}{\sqrt{-g}}D_\mu\!\left(\sqrt{-g}\,g^{\mu\nu}D_\nu\Psi\right) = \mu_s^2\Psi, \qquad D_\mu = \nabla_\mu - iqA_\mu,
\end{equation}
where $A_\mu$ is the electromagnetic four-potential of the BH given in Eq.~\eqref{EMpotential}. The presence of the gauge-covariant derivative $D_\mu$ distinguishes the present treatment from Ref.~\cite{Senjaya:2025kerremda}, where only neutral scalars ($q=0$) were studied; the coupling between $q$ and $A_\mu$ fundamentally modifies the singularity structure of the resulting ordinary differential equations (ODEs) and, consequently, all physical observables.

The Kerr-EMDA metric possesses two Killing vectors, $\partial_t$ and $\partial_\phi$, corresponding to stationarity and axial symmetry. Since $A_\mu$ also respects these symmetries, the KGE admits the separable ansatz
\begin{equation}\label{ansatz}
\Psi(t,r,\theta,\phi) = R(r)\,S(\theta)\,e^{im\phi}\,e^{-i\omega t},
\end{equation}
where $\omega\in\mathbb{C}$ is the frequency, $m\in\mathbb{Z}$ is the azimuthal quantum number, and $R(r)$, $S(\theta)$ are functions to be determined. Acting with the covariant derivatives on the ansatz, we obtain
\begin{equation}\label{Dcomponents}
D_t\Psi = -i\!\left(\omega + qA_t\right)\Psi, \qquad D_\phi\Psi = i\!\left(m - qA_\phi\right)\Psi, \qquad D_r\Psi = R'S\,e^{im\phi-i\omega t}, \qquad D_\theta\Psi = RS'\,e^{im\phi-i\omega t},
\end{equation}
where a prime denotes differentiation with respect to the argument.

To write the KGE explicitly, we need the inverse metric components. The metric determinant for the Kerr-EMDA line element~\eqref{metric} satisfies $\sqrt{-g} = \rho^2\sin\theta$, which has the same form as the Kerr case. Through the inversion of the $(t,\phi)$ block, whose determinant is $g_{tt}g_{\phi\phi} - g_{t\phi}^2 = -\Delta\sin^2\theta$, the non-vanishing contravariant components read
\begin{equation}\label{inversemetric}
g^{tt} = -\frac{A}{\rho^2\Delta}, \qquad g^{t\phi} = -\frac{rr_s\,a}{\rho^2\Delta}, \qquad g^{\phi\phi} = \frac{\Delta - a^2\sin^2\theta}{\rho^2\Delta\sin^2\theta}, \qquad g^{rr} = \frac{\Delta}{\rho^2}, \qquad g^{\theta\theta} = \frac{1}{\rho^2}.
\end{equation}
Substituting these into Eq.~\eqref{KGE} and inserting the ansatz~\eqref{ansatz}, the KGE separates into two pieces --- one depending solely on $r$ and the other solely on $\theta$ --- provided one defines the function
\begin{equation}\label{Kdef}
\mathcal{K}(r) \equiv \omega\!\left[r(r-2D)+a^2\right] - am + qQr.
\end{equation}
This function encapsulates the combined gravitational and electromagnetic coupling: the first term originates from $g^{tt}$ and $g^{t\phi}$ acting on $\omega$, the second from the frame-dragging interaction with $m$, and the third --- absent in Senjaya and Ponglertsakul's treatment --- from the scalar charge $q$ coupling to the BH charge $Q$.

\subsection{Separation of variables} \label{sec3b}

After substituting Eqs.~\eqref{ansatz}--\eqref{inversemetric} into the KGE, multiplying the resulting expression by $\rho^2$, and grouping the $r$- and $\theta$-dependent terms on opposite sides, we arrive at two separated ODEs linked by the separation constant $\lambda_{\ell m}$.

\medskip
\noindent\textbf{Radial equation.}
\begin{equation}\label{radialODE}
\frac{d}{dr}\!\left(\Delta\frac{dR}{dr}\right) + \left[\frac{\mathcal{K}^2}{\Delta} + 2am\omega - \lambda_{\ell m} - \mu_s^2\,r(r-2D) - a^2\omega^2\right]R = 0.
\end{equation}
This ODE has regular singular points at the two horizons $r = r_\pm$ (where $\Delta = 0$) and an irregular singular point of rank one at $r\to\infty$, placing it in the class of the confluent Heun equation (CHE).

\medskip
\noindent\textbf{Angular equation.}
\begin{equation}\label{angularODE}
\frac{1}{\sin\theta}\frac{d}{d\theta}\!\left(\sin\theta\frac{dS}{d\theta}\right) + \left[a^2(\omega^2 - \mu_s^2)\cos^2\theta - \frac{m^2}{\sin^2\theta} + \lambda_{\ell m}\right]S = 0.
\end{equation}
This is the well-known spheroidal harmonic equation. In the limit $a\omega\to 0$, the separation constant reduces to $\lambda_{\ell m}\to \ell(\ell+1)$ and $S(\theta)\to P_\ell^m(\cos\theta)$, where $P_\ell^m$ are the associated Legendre functions. We observe that the angular equation is identical to the one obtained in Ref.~\cite{Senjaya:2025kerremda}, since the gauge potential $A_\mu$ has no purely angular contribution --- the $\theta$-dependence of $A_t$ and $A_\phi$ factored out during the separation via the structure of $\rho^2$.

We note that two key differences between the present radial equation~\eqref{radialODE} and the one in Ref.~\cite{Senjaya:2025kerremda} arise from the $q\neq 0$ extension: (i) the function $\mathcal{K}(r)$ acquires the extra linear term $qQr$, which shifts the Frobenius indices at both horizons; and (ii) the superradiant threshold frequency, determined by the condition $\mathcal{K}(r_+)=0$, now involves the electrostatic potential of the BH.

\subsection{Angular solution: spheroidal harmonics} \label{sec3c}

To bring the angular equation~\eqref{angularODE} into the CHE form, we introduce the variable $y = (1 - \cos\theta)/2$, which maps $\theta = 0$ to $y = 0$ and $\theta = \pi$ to $y = 1$. Setting $c^2 = a^2(\omega^2-\mu_s^2)$ and writing $S(\theta)\to S_\ell^m(y)$, the angular equation becomes
\begin{equation}\label{angulary}
y(y-1)\frac{d^2S}{dy^2} + \left(y-\tfrac{1}{2}\right)\frac{dS}{dy} + \left[\frac{c^2(2y-1)^2}{4} + c^2 - \frac{m^2}{4y(1-y)} + \frac{\lambda_{\ell m}}{4}\right]S = 0.
\end{equation}
The singular points of this equation are $y=0$ and $y=1$ (both regular, corresponding to the north and south poles of the sphere) and $y\to\infty$ (irregular). Introducing the prefactor $S(y) = y^{|m|/2}(1-y)^{|m|/2}\,e^{-c\,(2y-1)/2}\,\mathcal{S}(y)$ absorbs the leading behavior at the poles and at infinity, reducing the equation for $\mathcal{S}(y)$ to the canonical CHE. The exact angular solution is therefore
\begin{equation}\label{angularsol}
S_\ell^m(y) = \mathcal{N}_S\,y^{|m|/2}(1-y)^{|m|/2}\,e^{-c\,(2y-1)/2}\;\text{HeunC}\!\left(\alpha_\theta,\,\beta_\theta,\,\gamma_\theta,\,\delta_\theta,\,\eta_\theta;\,y\right),
\end{equation}
where $\mathcal{N}_S$ is a normalization constant and the five angular Heun parameters are
\begin{equation}\label{angHeun}
\alpha_\theta = -2c, \qquad \beta_\theta = |m|, \qquad \gamma_\theta = |m|, \qquad \delta_\theta = -c^2 + c(|m|+1), \qquad \eta_\theta = \tfrac{1}{4}\!\left[\lambda_{\ell m} - (|m|+1)^2\right] + \tfrac{c}{2}(|m|+1).
\end{equation}
Imposing regularity at both poles ($y=0$ and $y=1$) quantizes $\lambda_{\ell m}$ for given $(\ell,m,c)$. In the massless limit $\mu_s\to 0$ with $a\omega\neq 0$, one has $c\to a\omega$ and recovers the standard spin-weighted spheroidal harmonics~\cite{Berti:2009kk}. The polynomial condition of the CHF (Appendix~\ref{appB}) applied to the angular equation gives the eigenvalue relation that determines $\lambda_{\ell m}$ as a function of $a\omega$ and $a\mu_s$.

\subsection{Radial equation: transformation to CHE form} \label{sec3d}

We now transform the radial equation~\eqref{radialODE} into the standard CHE. Since $\Delta = (r - r_+)(r-r_-)$, we introduce the dimensionless variable
\begin{equation}\label{zetadef}
\zeta = \frac{r - r_-}{r_+ - r_-} \equiv \frac{r - r_-}{\delta_r},
\end{equation}
which maps the inner horizon $r_-$ to $\zeta = 0$, the outer horizon $r_+$ to $\zeta = 1$, and spatial infinity $r\to\infty$ to $\zeta\to +\infty$. Under this transformation, $\Delta = \delta_r^2\,\zeta(\zeta - 1)$ and the radial equation~\eqref{radialODE} becomes
\begin{equation}\label{radzetaraw}
\frac{d}{d\zeta}\!\left[\zeta(\zeta-1)\frac{dR}{d\zeta}\right] + \left[\frac{\mathcal{K}^2(\zeta)}{\delta_r^2\,\zeta(\zeta-1)} + \mathcal{V}(\zeta)\right]R = 0,
\end{equation}
where $\mathcal{K}(\zeta) = \mathcal{K}(r_- + \delta_r\zeta)$ is a quadratic polynomial in $\zeta$ and $\mathcal{V}(\zeta)$ collects the remaining terms. The equation has regular singular points at $\zeta = 0$ and $\zeta=1$, and an irregular singular point at $\zeta\to\infty$.

The values of $\mathcal{K}$ at the horizons play a central role. Using $r_\pm(r_\pm - 2D)+a^2 = 2Mr_\pm$ (which follows from $\Delta(r_\pm) = 0$), we obtain
\begin{equation}\label{Khorizons}
\mathcal{K}_\pm \equiv \mathcal{K}(r_\pm) = (2M\omega + qQ)\,r_\pm - am.
\end{equation}
We can also express $\mathcal{K}_+$ in terms of the horizon quantities:
\begin{equation}\label{Kplus}
\mathcal{K}_+ = \Sigma_+\!\left(\omega - m\Omega_H + q\Phi_H\right) = 2Mr_+\!\left(\omega - m\Omega_H + q\Phi_H\right).
\end{equation}
This confirms that $\mathcal{K}_+$ vanishes precisely at the generalized superradiant bound $\omega_c = m\Omega_H - q\Phi_H$, where the sign of $q\Phi_H$ is determined by the convention $D_\mu = \nabla_\mu - iqA_\mu$ with $A_t < 0$.

\medskip
\noindent\textbf{Frobenius analysis.} Near $\zeta = 1$ (outer horizon), the leading behavior of the radial function is $R\sim (\zeta-1)^{s_+}$, where the indicial equation gives
\begin{equation}\label{indicial_plus}
s_+ = \pm\frac{i\mathcal{K}_+}{\delta_r} = \pm\frac{i\Sigma_+(\omega - m\Omega_H + q\Phi_H)}{2\sqrt{(M+D)^2 - a^2}}.
\end{equation}
The purely ingoing boundary condition at the event horizon requires the negative sign for $s_+$ (assuming $\omega > \omega_c$), giving $R\sim (\zeta-1)^{-i\mathcal{K}_+/\delta_r}$. In the tortoise coordinate $r_* \to -\infty$ at the horizon, this corresponds to $\Psi\sim e^{-i(\omega - m\Omega_H + q\Phi_H)v}$, where $v$ is the ingoing Eddington-Finkelstein coordinate. Similarly, near $\zeta = 0$ (inner horizon), $R\sim \zeta^{s_-}$ with $s_- = \pm i\mathcal{K}_-/\delta_r$.

Near $\zeta\to\infty$ (spatial infinity), the radial equation has an irregular singularity controlled by
\begin{equation}\label{asymptotic}
R\sim \zeta^p\,e^{\pm\sqrt{\mu_s^2-\omega^2}\,\delta_r\,\zeta}, \qquad p \in \mathbb{C},
\end{equation}
where the decaying solution (required for quasibound states) selects the sign with $\text{Re}[\sqrt{\mu_s^2-\omega^2}\,\delta_r\,\zeta]>0$.

\medskip
\noindent\textbf{Reduction to CHE.} To strip off the singular behavior, we write
\begin{equation}\label{Rsubstitution}
R(\zeta) = \zeta^{\tilde{\beta}/2}\,(\zeta - 1)^{\tilde{\gamma}/2}\,e^{\tilde{\alpha}\zeta/2}\,F(\zeta),
\end{equation}
where $\tilde{\alpha}$, $\tilde{\beta}$, $\tilde{\gamma}$ are chosen to absorb the leading singularities. The function $F(\zeta)$ then satisfies the CHE in canonical form~\cite{Ronveaux:1995,Fiziev:2009}
\begin{equation}\label{CHE}
F'' + \!\left(\tilde{\alpha} + \frac{1+\tilde{\beta}}{\zeta} + \frac{1+\tilde{\gamma}}{\zeta - 1}\right)\!F' + \frac{\tilde{\mu}\,\zeta - \tilde{q}_0}{\zeta(\zeta-1)}\,F = 0,
\end{equation}
where $\tilde{\mu}$ and $\tilde{q}_0$ are related to the Heun parameters $(\tilde{\delta},\tilde{\eta})$ through
\begin{equation}\label{muq0}
\tilde{\mu} = \frac{\tilde{\alpha}}{2}(\tilde{\beta}+\tilde{\gamma}+2) + \tilde{\delta}, \qquad \tilde{q}_0 = \frac{1+\tilde{\beta}}{2}\!\left(\frac{\tilde{\alpha}}{2} + 1 + \tilde{\gamma}\right) - \tilde{\eta} - \frac{\tilde{\delta}}{2}.
\end{equation}

\subsection{Heun parameters for the charged radial equation} \label{sec3e}

After a calculation that we detail in Appendix~\ref{appA}, we obtain the five CHE parameters for the radial equation:

\medskip
\noindent (i) The parameter controlling the irregular singularity at infinity:
\begin{equation}\label{alphatilde}
\tilde{\alpha} = -2i\delta_r\sqrt{\mu_s^2 - \omega^2}.
\end{equation}
For bound states ($\text{Re}[\omega]<\mu_s$), $\tilde{\alpha}$ has a negative real part, ensuring that $e^{\tilde{\alpha}\zeta/2}\to 0$ as $\zeta\to\infty$.

\medskip
\noindent (ii) The Frobenius exponent at the inner horizon ($\zeta=0$):
\begin{equation}\label{betatilde}
\tilde{\beta} = \frac{2i\mathcal{K}_-}{\delta_r} = \frac{2i\!\left[(2M\omega+qQ)\,r_- - am\right]}{\delta_r}.
\end{equation}

\medskip
\noindent (iii) The Frobenius exponent at the outer horizon ($\zeta=1$):
\begin{equation}\label{gammatilde}
\tilde{\gamma} = -\frac{2i\mathcal{K}_+}{\delta_r} = -\frac{2i\!\left[(2M\omega+qQ)\,r_+ - am\right]}{\delta_r} = -\frac{2i\Sigma_+(\omega - m\Omega_H + q\Phi_H)}{\delta_r}.
\end{equation}
The sign is chosen so that $(\zeta-1)^{\tilde{\gamma}/2}$ represents the purely ingoing wave at the outer horizon.

\medskip
\noindent (iv) The parameter $\tilde{\delta}$:
\begin{equation}\label{deltatilde}
\tilde{\delta} = i\delta_r\!\left(2\omega^2 - \mu_s^2\right)\frac{1}{\sqrt{\mu_s^2 - \omega^2}} - 2i\delta_r\sqrt{\mu_s^2-\omega^2}\!\left(M+D\right) + i\frac{2am\omega - \lambda_{\ell m} - a^2\omega^2}{\delta_r\sqrt{\mu_s^2-\omega^2}}\,.
\end{equation}

\medskip
\noindent (v) The accessory parameter $\tilde{\eta}$:
\begin{equation}\label{etatilde}
\tilde{\eta} = \frac{1}{2}(1+\tilde{\beta})(1+\tilde{\gamma}) + \frac{1}{4}\tilde{\alpha}(1+\tilde{\beta}) - \frac{\tilde{\delta}}{2} + \frac{2am\omega - \lambda_{\ell m} - a^2\omega^2 - \mu_s^2\,r_-(r_- - 2D)}{2} + \frac{\mathcal{K}_-^2}{\delta_r^2}\,.
\end{equation}

Several observations are in order. First, $\tilde{\beta}$ and $\tilde{\gamma}$ are both purely imaginary (for real $\omega$), reflecting the oscillatory nature of the wave function at the horizons. Second, the ratio $\tilde{\gamma}/(-2i)$ equals $(\omega - m\Omega_H + q\Phi_H)/\kappa_s$, a combination that will appear naturally in the Hawking thermal factor (Sec.~\ref{sec7}). Third, in the limit $q\to 0$, the parameters reduce to those of the neutral scalar case studied in Ref.~\cite{Senjaya:2025kerremda}, since $\mathcal{K}_\pm \to 2M\omega\,r_\pm - am$. Fourth, in the further limit $D\to 0$, we recover the known Kerr results~\cite{Vieira:2023EGB,Fiziev:2009}. Finally, the five parameters satisfy the consistency relation
\begin{equation}\label{paramcheck}
\tilde{\beta} + \tilde{\gamma} = -\frac{2i}{\delta_r}\!\left[\mathcal{K}_+ - \mathcal{K}_-\right] = -\frac{2i}{\delta_r}\!\left(2M\omega + qQ\right)\!\left(r_+ - r_-\right) = -2i(2M\omega + qQ),
\end{equation}
which provides a non-trivial algebraic check on the derivation. In Table~\ref{tab:Heun}, we list the Heun parameters for several representative BH configurations with $M=1$, $m=1$, $\omega = 0.4\,M^{-1}$, and $\mu_s = 0.5\,M^{-1}$.

\begin{table*}[htbp]
\centering
\caption{CHE parameters for charged massive KGE on Kerr-EMDA. All quantities are computed for $M=1$, $m=1$, $\omega = 0.4\,M^{-1}$, and $\mu_s = 0.5\,M^{-1}$. The columns show the BH spin $a$, charge $Q$, scalar charge $q$, dilaton parameter $D$, horizon radii $r_\pm$, superradiant bound $\omega_c = m\Omega_H - q\Phi_H$, and imaginary parts of $\tilde{\beta}$ and $\tilde{\gamma}$ (both are purely imaginary for real $\omega$). The Kerr and neutral-scalar limits are included for comparison.}
\label{tab:Heun}
\footnotesize
\setlength{\tabcolsep}{6pt}
\begin{tabular}{l c c c c c c c c c}
\hline\hline
Configuration & $a/M$ & $Q/M$ & $q$ & $D/M$ & $r_+/M$ & $r_-/M$ & $\omega_c M$ & Im$(\tilde{\beta})$ & Im$(\tilde{\gamma})$ \\
\hline
Kerr ($D\!=\!Q\!=\!q\!=\!0$)  & 0.5 & 0.0 & 0.0 & 0.0000 & 1.8660 & 0.1340 & 0.1340 & $-0.4536$ & $-1.1464$ \\
Neutral ($q\!=\!0$)           & 0.5 & 0.3 & 0.0 & 0.0450 & 1.9626 & 0.1274 & 0.1274 & $-0.4338$ & $-1.1662$ \\
$q\!=\!0.2$                  & 0.5 & 0.3 & 0.2 & 0.0450 & 1.9626 & 0.1274 & 0.0974 & $-0.4255$ & $-1.2945$ \\
$q\!=\!0.5$                  & 0.5 & 0.3 & 0.5 & 0.0450 & 1.9626 & 0.1274 & 0.0524 & $-0.4130$ & $-1.4870$ \\
$q\!=\!1.0$                  & 0.5 & 0.3 & 1.0 & 0.0450 & 1.9626 & 0.1274 & $-0.0226$ & $-0.3922$ & $-1.8078$ \\
$Q\!=\!0.6M$                 & 0.5 & 0.6 & 0.2 & 0.1800 & 2.2488 & 0.1112 & 0.0512 & $-0.3721$ & $-1.4679$ \\
$a\!=\!0.3M$                 & 0.3 & 0.3 & 0.2 & 0.0450 & 2.0460 & 0.0440 & 0.0433 & $-0.2619$ & $-1.4581$ \\
$a\!=\!0.7M$                 & 0.7 & 0.3 & 0.2 & 0.0450 & 1.8209 & 0.2691 & 0.1622 & $-0.6039$ & $-1.1161$ \\
\hline\hline
\end{tabular}
\end{table*}

Table~\ref{tab:Heun} reveals several physically important trends. Increasing $q$ at fixed $(a,Q)$ shifts the superradiant bound $\omega_c$ to lower values: for $q = 1.0$, the bound becomes negative, meaning that the superradiant condition $\omega < \omega_c$ cannot be satisfied for positive real frequencies. This signals a qualitative change in the scattering physics of strongly charged scalars. The Frobenius index Im$(\tilde{\gamma})$ at the outer horizon grows in magnitude with $q$, reflecting the enhanced coupling between the scalar and BH charges. Conversely, Im$(\tilde{\beta})$ at the inner horizon is less sensitive to $q$, since the inner horizon radius $r_-$ is much smaller than $r_+$ for the sub-extremal configurations considered.

The behavior of $\mathcal{K}(r)$ and the superradiant condition are illustrated in Fig.~\ref{fig:K_superrad}. Panel (a) shows that $\mathcal{K}(r)$ is an upward-opening quadratic that converges to the common value $\mathcal{K}(0) = a^2\omega - am \simeq -0.4$ at $r=0$ (independent of $q$), with increasing $q$ lifting the curves at large $r$ through the $qQr$ term. The dashed line at $r_+ \simeq 1.96M$ shows that $\mathcal{K}(r_+)$ ranges from $0.89$ (neutral case) to $1.77$ ($q=1.0$). Panel (b) confirms the linear dependence $\omega_c = m\Omega_H - q\Phi_H$, with each spin value yielding a straight line whose slope $-\Phi_H$ is steeper for smaller $a$ (where $\Phi_H$ is larger). The zero-crossing at $q_{\text{cr}} = m\Omega_H/\Phi_H$ sets the boundary beyond which superradiant amplification is forbidden.

\begin{figure}[htbp]
\centering
\begin{subfigure}[t]{0.48\textwidth}
\centering
\includegraphics[width=\textwidth]{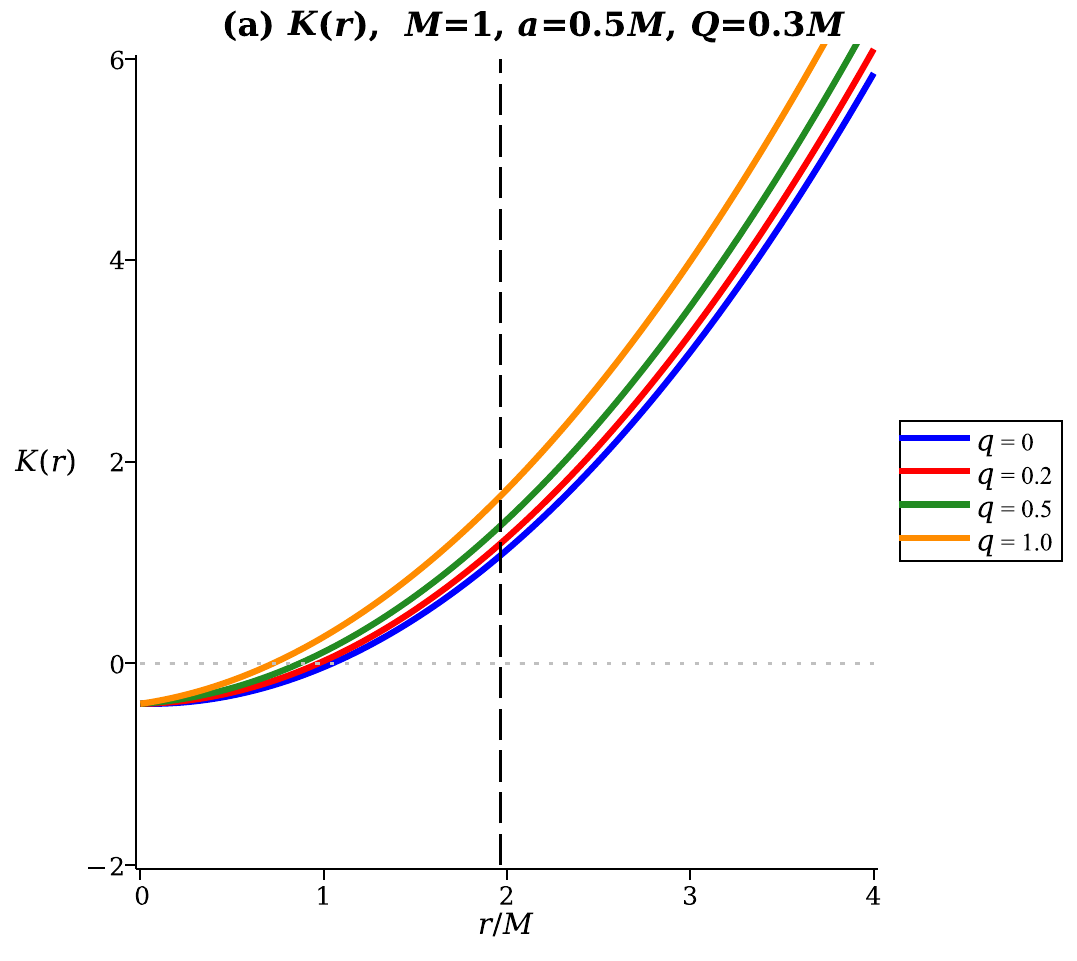}
\end{subfigure}
\hfill
\begin{subfigure}[t]{0.48\textwidth}
\centering
\includegraphics[width=\textwidth]{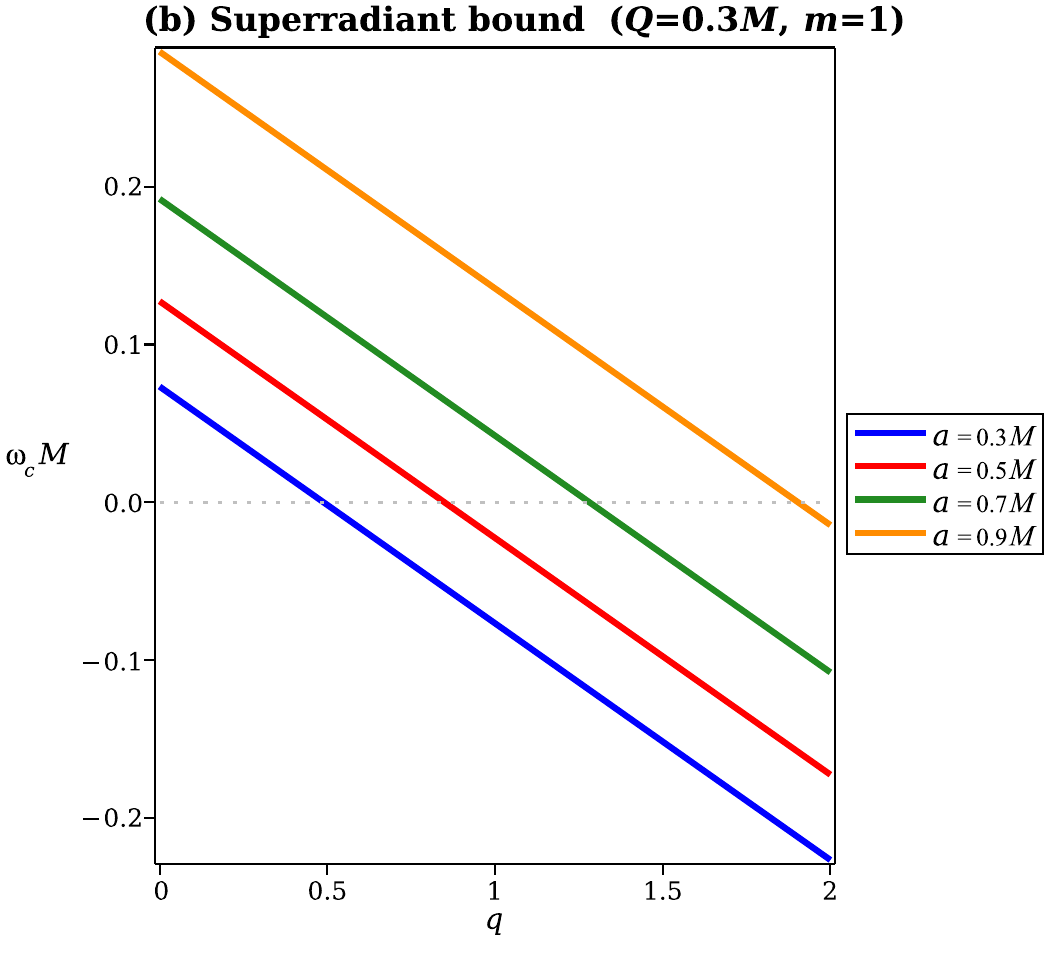}
\end{subfigure}
\caption{(a) The function $\mathcal{K}(r) = \omega\Sigma_r - am + qQr$ for several values of the scalar charge $q$, with $M=1$, $a=0.5M$, $Q=0.3M$, $m=1$, and $\omega=0.4\,M^{-1}$. The dashed vertical line marks the outer horizon $r_+ \simeq 1.96M$. Increasing $q$ raises $\mathcal{K}$ due to the positive $qQr$ contribution, with the curves merging near $r \approx 0$ where $\mathcal{K} \to a^2\omega - am \simeq -0.4$. (b) The superradiant bound $\omega_c = m\Omega_H - q\Phi_H$ as a function of $q$ for different BH spins at fixed $Q=0.3M$ and $m=1$. Each curve is linear with slope $-\Phi_H(a)$. For small $a$ (e.g.\ $a=0.3M$), $\omega_c$ crosses zero at lower $q$, quenching superradiance. Higher spins extend the superradiant window to larger $q$.}
\label{fig:K_superrad}
\end{figure}

\subsection{Complete exact solution and boundary conditions} \label{sec3f}

The general solution of the CHE~\eqref{CHE} is a linear combination of two linearly independent solutions, each regular at $\zeta = 0$. Reverting to the original radial function via Eq.~\eqref{Rsubstitution}, we obtain
\begin{equation}\label{radialsol}
R(\zeta) = \zeta^{\tilde{\beta}/2}\,(\zeta - 1)^{\tilde{\gamma}/2}\,e^{\tilde{\alpha}\zeta/2}\!\left[C_1\,\text{HeunC}(\tilde{\alpha},\tilde{\beta},\tilde{\gamma},\tilde{\delta},\tilde{\eta};\,\zeta) + C_2\,\zeta^{-\tilde{\beta}}\,\text{HeunC}(\tilde{\alpha},-\tilde{\beta},\tilde{\gamma},\tilde{\delta},\tilde{\eta};\,\zeta)\right],
\end{equation}
where $C_1$ and $C_2$ are constants. The first solution, with prefactor $\zeta^{\tilde{\beta}/2}$, and the second, with prefactor $\zeta^{-\tilde{\beta}/2}$ (since the $\zeta^{-\tilde{\beta}}$ from the second HeunC combines with the $\zeta^{\tilde{\beta}/2}$ prefactor), correspond to the two Frobenius roots at the inner horizon.

\medskip
\noindent\textbf{Boundary conditions.} We impose two physical requirements:

\medskip
\noindent(a) \emph{Purely ingoing wave at the outer horizon} ($\zeta\to 1^+$). Near the event horizon, the radial function must behave as $R\sim (\zeta-1)^{\tilde{\gamma}/2}$, which translates to $R\sim (r - r_+)^{-i(\omega - m\Omega_H + q\Phi_H)/(2\kappa_s)}$ in the original coordinate. This ensures that no outgoing flux crosses the horizon. Both linearly independent solutions in Eq.~\eqref{radialsol} satisfy this condition automatically through the common prefactor $(\zeta-1)^{\tilde{\gamma}/2}$.

\medskip
\noindent(b) \emph{Decay at spatial infinity} ($\zeta\to\infty$). For quasibound states, we require $|R|\to 0$ as $r\to\infty$. The asymptotic behavior of $\text{HeunC}$ as $\zeta\to\infty$ involves $e^{-\tilde{\alpha}\zeta}$~\cite{Ronveaux:1995}, so that $R\sim e^{\tilde{\alpha}\zeta/2}\cdot e^{-\tilde{\alpha}\zeta} = e^{-\tilde{\alpha}\zeta/2}$. Since $\text{Re}[\tilde{\alpha}/2] < 0$ for bound states, the overall solution $R\sim e^{(\tilde{\alpha}-\tilde{\alpha})\zeta/2}$ requires care: one must truncate the HeunC to a polynomial so that the $e^{-\tilde{\alpha}\zeta}$ growth does not appear. This truncation is achieved by the polynomial condition of the CHF, discussed in Sec.~\ref{sec4}.

Combining Eqs.~\eqref{ansatz}, \eqref{angularsol}, and~\eqref{radialsol}, the complete exact wave function for a charged massive scalar field on the Kerr-EMDA background is
\begin{equation}\label{fullPsi}
{\Psi(t,r,\theta,\phi) = \mathcal{N}\,R(\zeta)\,S_\ell^m(y)\,e^{im\phi}\,e^{-i\omega t},}
\end{equation}
where $\mathcal{N}$ is an overall normalization constant, $\zeta = (r-r_-)/\delta_r$, and $y = (1-\cos\theta)/2$. Both $R$ and $S_\ell^m$ are expressed entirely in terms of confluent Heun functions (CHFs), and the full solution is parametrized by the quantum numbers $(\ell, m, n)$, the BH parameters $(M, a, Q)$, and the field parameters $(\mu_s, q, \omega)$. The frequency $\omega$ is not free but is determined by the boundary conditions; its explicit form is derived in the next section.

We remark that the separability of the charged KGE on the Kerr-EMDA background is not obvious a priori: the gauge potential components $A_t$ and $A_\phi$ both depend on $\theta$ through $\rho^2$ [see Eq.~\eqref{EMpotential}]. The separation succeeds because the $\theta$-dependent parts of $qA_t$ and $qA_\phi$ cancel against corresponding terms from the inverse metric when the equation is multiplied by $\rho^2$. What remains is the purely radial function $\mathcal{K}(r)$ defined in Eq.~\eqref{Kdef}, which absorbs all gauge contributions into the radial sector. This cancellation is a direct consequence of the Kerr-like structure of the Kerr-EMDA metric and holds for the same reason that the neutral KGE separates on Kerr-Newman backgrounds~\cite{Brito:2015oca}.

\section{Resonant Frequencies and Entropy Quantization} \label{sec4}

Having established the exact radial solution in terms of CHFs in Sec.~\ref{sec3}, we now extract the physical consequences of the polynomial truncation condition. The requirement that the CHF reduces to a polynomial of finite degree selects a discrete, equispaced tower of complex resonant frequencies $\omega_n$, whose imaginary-part spacing $|\Delta\omega_I| = 1/(2M)$ is entirely fixed by the BH mass and is independent of the scalar charge $q$, field mass $\mu_s$, angular momentum quantum numbers $(m,\ell)$, and the BH spin $a$ and charge $Q$. Through the Maggiore prescription and the first law of BH thermodynamics, this spacing translates into a quantized entropy spectrum $\delta S_{\text{BH}} = 4\pi r_+/(r_+ - r_-)$ that reduces to $4\pi$ for Schwarzschild but depends on the BH parameters in general, diverging at extremality. This section is organized as follows: Sec.~\ref{sec4a} derives the resonant frequency spectrum, Sec.~\ref{sec4b} obtains the entropy and area quantization, and Sec.~\ref{sec4c} discusses the highly-damped regime and its connection to Hod's conjecture.

\subsection{Resonant frequency spectrum from the $\varepsilon_n$-condition} \label{sec4a}

The CHF polynomial condition (see Appendix~\ref{appB}) stipulates that the confluent Heun function $\text{HeunC}(\tilde{\alpha},\tilde{\beta},\tilde{\gamma},\tilde{\delta},\tilde{\eta};\,\zeta)$ reduces to a polynomial of degree $n$ (with $n = 0,1,2,\ldots$) in $\zeta$ when the parameters satisfy
\begin{equation}\label{epsiloncondition}
\varepsilon_n \equiv \frac{\tilde{\delta}}{\tilde{\alpha}} + \frac{\tilde{\beta}+\tilde{\gamma}}{2} + 1 = -n, \qquad n = 0,\,1,\,2,\,\ldots
\end{equation}
This polynomial truncation simultaneously enforces the decaying boundary condition at spatial infinity and the ingoing condition at the horizon, thereby selecting a discrete set of complex frequencies $\omega_n$. In what follows, we substitute the Heun parameters from Sec.~\ref{sec3e} into Eq.~\eqref{epsiloncondition} and solve for $\omega$.

From Eqs.~\eqref{alphatilde} and~\eqref{deltatilde}, the ratio $\tilde{\delta}/\tilde{\alpha}$ reads
\begin{equation}\label{deltaalpha}
\frac{\tilde{\delta}}{\tilde{\alpha}} = -\frac{2\omega^2 - \mu_s^2}{2(\mu_s^2 - \omega^2)} + (M+D) - \frac{2am\omega - \lambda_{\ell m} - a^2\omega^2}{2\delta_r^2(\mu_s^2 - \omega^2)}\,.
\end{equation}
We simplify the first term using the identity
\begin{equation}\label{identity1}
\frac{2\omega^2 - \mu_s^2}{2(\mu_s^2 - \omega^2)} = -1 + \frac{\mu_s^2}{2(\mu_s^2 - \omega^2)}\,,
\end{equation}
so that
\begin{equation}\label{deltaalpha2}
\frac{\tilde{\delta}}{\tilde{\alpha}} = 1 - \frac{\mu_s^2}{2(\mu_s^2 - \omega^2)} + (M+D) - \frac{2am\omega - \lambda_{\ell m} - a^2\omega^2}{2\delta_r^2(\mu_s^2 - \omega^2)}\,.
\end{equation}
From the consistency relation~\eqref{paramcheck}, we have $(\tilde{\beta}+\tilde{\gamma})/2 = -i(2M\omega + qQ)$. Inserting these into Eq.~\eqref{epsiloncondition} and rearranging:
\begin{equation}\label{omega_implicit}
-i(2M\omega + qQ) + (M+D) + 2 - \frac{\mu_s^2}{2(\mu_s^2 - \omega^2)} - \frac{2am\omega - \lambda_{\ell m} - a^2\omega^2}{2\delta_r^2(\mu_s^2 - \omega^2)} = -n.
\end{equation}

{\color{black}We introduce the notation
$\varpi \equiv \sqrt{\mu_s^2 - \omega^2}$
(with $\text{Re}[\varpi]>0$ for bound states) and define the
``orbital'' contribution}
\begin{equation}\label{Ldef}
{\color{black}\mathcal{L} \equiv \frac{\mu_s^2}{2\varpi^2} +
\frac{2am\omega - \lambda_{\ell m} - a^2\omega^2}{2\delta_r^2\varpi^2}
= \frac{\mu_s^2\delta_r^2 + 2am\omega - \lambda_{\ell m} -
a^2\omega^2}{2\delta_r^2\varpi^2}\,.}
\end{equation}
Then Eq.~\eqref{omega_implicit} takes the compact form
\begin{equation}\label{omega_compact}
-i(2M\omega + qQ) + (M+D) + 2 - \mathcal{L} = -n.
\end{equation}
Solving for $\omega$ (multiply both sides by $i$: $2M\omega + qQ = i[\mathcal{L} - n - (M+D) - 2]$):
\begin{equation}\label{omegan_general}
\omega_n = \frac{i}{2M}\!\left[\mathcal{L} - n - (M+D) - 2\right] - \frac{qQ}{2M}\,.
\end{equation}
This is an implicit equation since $\mathcal{L}$ depends on $\omega$ \textcolor{black}{through $\varpi^2 = \mu_s^2 - \omega^2$.} Nevertheless, Eq.~\eqref{omegan_general} admits a closed-form solution order by order in $1/n$ for highly damped modes and in $a\omega$ for slowly rotating BHs. Before analyzing these limits, we extract the general structure of the spectrum.

\medskip
\noindent\textbf{Real and imaginary parts.} Writing $\omega_n = \omega_R^{(n)} + i\,\omega_I^{(n)}$, Eq.~\eqref{omegan_general} yields:
\begin{equation}\label{omegaI}
\omega_I^{(n)} = \frac{\text{Re}[\mathcal{L}] - n - (M+D) - 2}{2M} \xrightarrow{n\gg 1} -\frac{n}{2M}\,,
\end{equation}
\begin{equation}\label{omegaR}
\omega_R^{(n)} = -\frac{qQ}{2M} - \frac{\text{Im}[\mathcal{L}]}{2M}\,.
\end{equation}
{\color{black}The imaginary part is negative and its magnitude grows linearly with the
overtone number $n$, signalling that higher overtones are progressively more damped ---
a hallmark of quasibound states. Taking the difference between consecutive modes from
Eq.~\eqref{omegaI}, the spacing of the imaginary parts reads
\begin{equation}\label{spacing}
\Delta\omega_I \equiv \omega_I^{(n+1)} - \omega_I^{(n)}
= -\frac{1}{2M}
+ \frac{\text{Re}[\mathcal{L}(\omega_{n+1})]
      - \text{Re}[\mathcal{L}(\omega_n)]}{2M}\,.
\end{equation}
The first term originates from the integer step $n\to n+1$ in the $\varepsilon_n$-condition
and depends only on the BH mass. The second term is a residual correction arising because
$\mathcal{L}$ [Eq.~\eqref{Ldef}] depends on $\omega$ through
$\Gamma^2 = \mu_s^2 - \omega^2$, so its value shifts from one overtone to the next.
In the highly-damped regime ($n\gg 1$), $|\omega|^2 \gg \mu_s^2$ drives
$\mathcal{L}\to -1/2 + \mathcal{O}(1/n)$; consequently,
$\text{Re}[\mathcal{L}(\omega_{n+1})] - \text{Re}[\mathcal{L}(\omega_n)] \to 0$
and the spacing reduces to
\begin{equation}\label{spacing_exact}
|\Delta\omega_I| = \frac{1}{2M}\,,\qquad n\gg 1\,.
\end{equation}
This asymptotic spacing is independent of $q$, $\mu_s$, $a$, $m$, and $\ell$; it is
determined solely by the BH mass. Table~\ref{tab:resonant} confirms the rapid convergence:
$|\Delta\omega_I|$ already reaches 99.6\% of the asymptotic value at $n = 5$. Since the
Maggiore prescription for entropy quantization [Eq.~\eqref{physfreq}] is formulated in the
highly-damped limit, the spacing $|\Delta\omega_I| = 1/(2M)$ enters the entropy
quantum~\eqref{deltaSBH} without approximation.}

{\color{black}It is useful to express this asymptotic spacing in terms of the surface gravity.
Using $\kappa_s = (r_+ - r_-)/(2\Sigma_+) = \delta_r/(2\Sigma_+)$ together with the
identity $\Sigma_+ = 2Mr_+$ [which follows from $r_+(r_+ - 2D) + a^2 = 2Mr_+$ via
$\Delta(r_+) = 0$; see the supplementary Maple file for the symbolic proof], one finds
\begin{equation}\label{spacing_kappa}
|\Delta\omega_I| = \frac{1}{2M} = \kappa_s \cdot \frac{r_+}{\delta_r}\,.
\end{equation}
Since $r_+/\delta_r > 1$ for any sub-extremal Kerr-EMDA BH with two distinct horizons,
the resonant frequency spacing exceeds the surface gravity:
$|\Delta\omega_I| > \kappa_s$ in general. The two coincide only when $r_+ = \delta_r$,
i.e.\ $r_- = 0$, which holds in the Schwarzschild limit ($a = D = 0$) and in the static
GMGHS case ($a = 0$).}

This structure parallels the RLDBH result of Ref.~\cite{Sakalli:2016mnk,Sakalli:2017}, where the resonant frequencies take the form $\omega_n = m\Omega_H + q\Phi_e + i(n+1)\kappa_s$, with the imaginary-part spacing equaling the surface gravity. In the Kerr-EMDA case, the $\varepsilon_n$-condition from the exact CHF polynomial truncation gives $|\Delta\omega_I| = 1/(2M)$ rather than $\kappa_s$. The two coincide in the Schwarzschild limit ($a=0$, $D=0$), where $r_+ = 2M$, $\delta_r = 2M$, and $\kappa_s = 1/(4M)$, so $|\Delta\omega_I| = 1/(2M) = 2\kappa_s$. We return to this comparison in Sec.~\ref{sec8}.

\subsection{Entropy and area quantization} \label{sec4b}

The equidistant spacing of the imaginary part of $\omega_n$ has direct implications for BH thermodynamics via Bekenstein's area quantization proposal~\cite{Bekenstein:1974jk}. The central argument, refined by Kunstatter~\cite{Kunstatter:2002} and Maggiore~\cite{Maggiore:2007nq}, proceeds as follows.

Consider a BH in a quasibound state with frequency $\omega_n$. The transition $n \to n+1$ changes the BH energy (mass) by an amount
\begin{equation}\label{dE}
\delta E = \hbar\,\Delta\omega_R + i\hbar\,\Delta\omega_I.
\end{equation}
In the Maggiore prescription~\cite{Maggiore:2007nq}, the physically relevant transition frequency is the magnitude
\begin{equation}\label{Maggiore}
\Delta\omega_{\text{phys}} = \sqrt{(\Delta\omega_R)^2 + (\Delta\omega_I)^2}\,.
\end{equation}
For highly damped modes ($n\gg 1$), $|\Delta\omega_I| \gg |\Delta\omega_R|$ since $|\Delta\omega_I| = 1/(2M)$ is fixed while $\Delta\omega_R$ depends on the slowly varying orbital contribution $\mathcal{L}$. Therefore,
\begin{equation}\label{physfreq}
\Delta\omega_{\text{phys}} \approx |\Delta\omega_I| = \frac{1}{2M}\,.
\end{equation}

The first law of BH thermodynamics applied to a stationary perturbation that changes only the mass (at fixed $J$ and $Q$) gives
\begin{equation}\label{firstlawperturbation}
\delta M = T_H\,\delta S_{\text{BH}} = \frac{\kappa_s}{2\pi}\,\delta S_{\text{BH}}\,.
\end{equation}
Identifying $\delta M = \hbar\,\Delta\omega_{\text{phys}}$ and setting $\hbar = 1$:
\begin{equation}\label{deltaSBH}
\delta S_{\text{BH}} = \frac{2\pi\,\Delta\omega_{\text{phys}}}{\kappa_s} = \frac{2\pi}{2M\kappa_s} = \frac{2\pi\cdot 2\Sigma_+}{2M\delta_r} = \frac{2\pi\cdot 2\cdot 2Mr_+}{2M\delta_r} = \frac{4\pi r_+}{\delta_r}\,.
\end{equation}
This expression is \emph{not} simply $2\pi$ in general; rather, it depends on the ratio $r_+/\delta_r$, which encodes the BH parameters. We can write it as
\begin{equation}\label{deltaSBH2}
{\delta S_{\text{BH}} = \frac{4\pi r_+}{r_+ - r_-} = \frac{4\pi(M+D+\sqrt{(M+D)^2-a^2})}{2\sqrt{(M+D)^2-a^2}}\,.}
\end{equation}

Let us examine the limiting cases:

\medskip
\noindent(i) \emph{Schwarzschild limit} ($a=0$, $D=0$): $r_+ = 2M$, $r_- = 0$, so $\delta S_{\text{BH}} = 4\pi\cdot 2M/(2M) = 4\pi$. The corresponding area quantum is $\delta\mathcal{A} = 4\cdot\delta S_{\text{BH}} = 16\pi\ell_P^2$, which matches the result obtained from the Schwarzschild QNM analysis~\cite{Maggiore:2007nq}.

\medskip
\noindent(ii) \emph{Slowly rotating limit} ($a\ll M+D$): Expanding to leading order in $a$,
\begin{equation}\label{slowrot}
\delta S_{\text{BH}} \approx 4\pi + \frac{\pi a^2}{(M+D)^2} + \frac{3\pi a^4}{4(M+D)^4} + \mathcal{O}(a^6).
\end{equation}

\medskip
\noindent(iii) \emph{Near-extremal limit} ($a\to (M+D)$): As $\delta_r \to 0$, $\delta S_{\text{BH}}\to\infty$, indicating that the entropy quantum diverges at extremality. This is consistent with the vanishing of $T_H$ and $\kappa_s$ in the extremal limit: an infinite number of quanta are required to change the entropy by a finite amount. {\color{black}The entropy-function approach for rotating extremal
solutions with axionic hair, including the connection to
Kerr-Sen-type attractors, has been developed by
dos~Santos~Costa~Filho~\cite{dosSantosCostaFilho:2025ibq}; a
microscopic understanding of the divergent entropy quantum
$\delta S_{\text{BH}} \to \infty$ at extremality may require input
from such near-horizon methods.}

\medskip
\noindent(iv) \emph{RLDBH comparison}: In the RLDBH of Ref.~\cite{Sakalli:2016mnk}, $\Delta\omega_I = \kappa_s$ exactly, which yields the universal $\delta S_{\text{BH}} = 2\pi$. The difference arises because the RLDBH metric has a different singularity structure (a single horizon at $r_+ = r_s$ with $r_- = 0$ effectively), whereas the Kerr-EMDA BH has two distinct horizons with a finite gap $\delta_r$.

The behavior of $\delta S_{\text{BH}}$ as a function of the spin parameter is shown in Fig.~\ref{fig:entropy_quant} and tabulated in Table~\ref{tab:entropy} for several BH configurations.

\begin{table}[htbp]
\centering
\caption{Entropy quantum $\delta S_{\text{BH}}$ and area quantum $\delta\mathcal{A}_H = 4\delta S_{\text{BH}}\,\ell_P^2$ for representative Kerr-EMDA configurations with $M=1$. The resonant frequency spacing is $|\Delta\omega_I| = 1/(2M) = 0.5$ for all entries. The Schwarzschild and Kerr limits are included for comparison.}
\label{tab:entropy}
\begin{tabular}{l c c c c c c c}
\hline\hline
Configuration & $a/M$ & $Q/M$ & $D/M$ & $r_+/M$ & $\delta_r/M$ & $\delta S_{\text{BH}}$ & $\delta\mathcal{A}_H/\ell_P^2$ \\
\hline
Schwarzschild & 0.0 & 0.0 & 0.000 & 2.0000 & 2.0000 & $4\pi$ & $16\pi$ \\
Kerr $a\!=\!0.5M$ & 0.5 & 0.0 & 0.000 & 1.8660 & 1.7321 & $4.309\pi$ & $17.24\pi$ \\
EMDA $Q\!=\!0.3M$ & 0.5 & 0.3 & 0.045 & 1.9626 & 1.8352 & $4.278\pi$ & $17.11\pi$ \\
EMDA $Q\!=\!0.6M$ & 0.5 & 0.6 & 0.180 & 2.2488 & 2.1377 & $4.208\pi$ & $16.83\pi$ \\
EMDA $a\!=\!0.7M$ & 0.7 & 0.3 & 0.045 & 1.8209 & 1.5518 & $4.694\pi$ & $18.78\pi$ \\
Near-extremal & 0.99 & 0.3 & 0.045 & 1.3796 & 0.6691 & $8.247\pi$ & $32.99\pi$ \\
\hline\hline
\end{tabular}
\end{table}

\begin{figure}[htbp]
\centering
\begin{subfigure}[t]{0.48\textwidth}
\centering
\includegraphics[width=\textwidth]{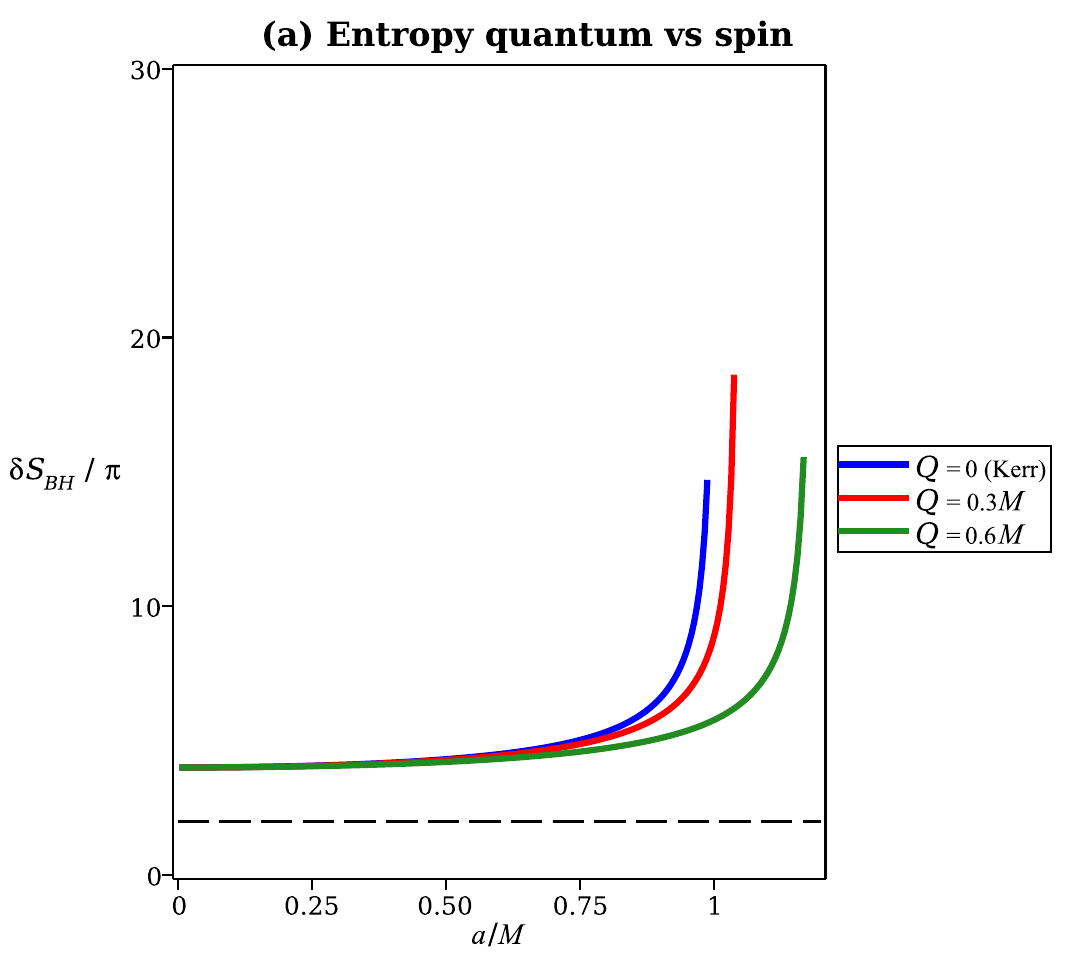}
\end{subfigure}
\hfill
\begin{subfigure}[t]{0.48\textwidth}
\centering
\includegraphics[width=\textwidth]{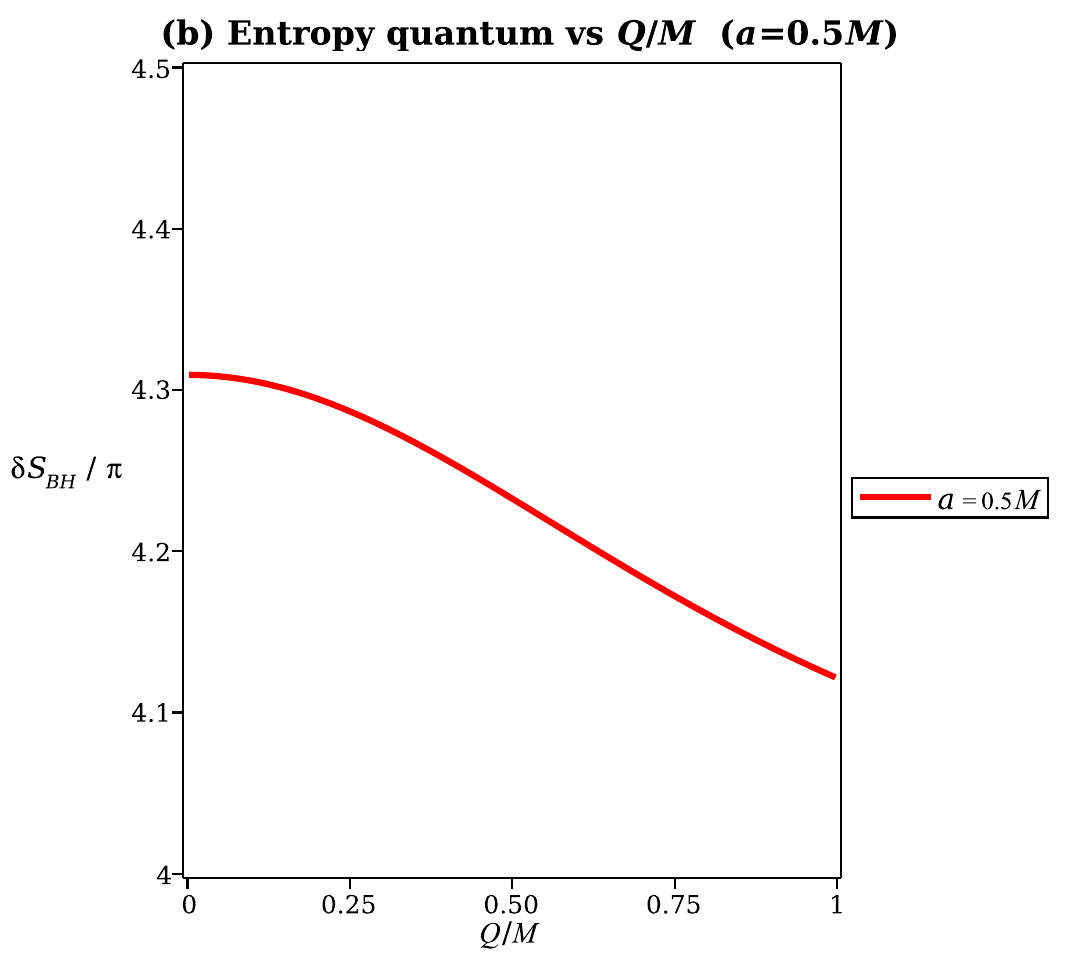}
\end{subfigure}
\caption{(a) Entropy quantum $\delta S_{\text{BH}}/\pi$ as a function of the spin parameter $a/M$ for $Q=0$ (Kerr, black), $Q=0.3M$ (red), and $Q=0.6M$ (green). The horizontal dashed line marks $\delta S_{\text{BH}} = 2\pi$. All curves start near $4\pi$ (Schwarzschild value) at $a=0$ and diverge at the respective extremal limits $a = M + D$, where $\delta_r \to 0$ and $T_H \to 0$. The dilaton shift $D = Q^2/(2M)$ extends the extremal spin beyond $M$ (e.g.\ $a_{\max} = 1.045M$ for $Q = 0.3M$ and $1.18M$ for $Q=0.6M$), broadening the sub-extremal domain. (b) Entropy quantum $\delta S_{\text{BH}}/\pi$ as a function of $Q/M$ at fixed spin $a = 0.5M$. The curve decreases monotonically from $\simeq 4.31$ (Kerr value at $Q=0$) to $\simeq 4.12$ at $Q = M$, reflecting the fact that the dilaton parameter $D = Q^2/(2M)$ increases $r_+$ and $\delta_r$ simultaneously, with $\delta_r$ growing faster and thereby reducing $\delta S_{\text{BH}} = 4\pi r_+/\delta_r$.}
\label{fig:entropy_quant}
\end{figure}

\subsection{Highly-damped regime and Hod's conjecture} \label{sec4c}

In the highly-damped regime $n\gg 1$, the quasibound state frequencies~\eqref{omegan_general} simplify considerably. Since $|\omega_I^{(n)}|\sim n/(2M)\to\infty$, we have $|\omega|^2 \gg \mu_s^2$, so {\color{black}$\varpi^2 = \mu_s^2 - \omega^2 \approx -\omega^2$} and consequently $\mathcal{L}\to -1/2 + \mathcal{O}(1/\omega)$. The asymptotic expansion of the resonant frequencies becomes
\begin{equation}\label{asymptotic_freq}
\omega_n \approx -\frac{qQ}{2M} - \frac{i}{2M}\!\left(n + M + D + \tfrac{5}{2}\right) + \mathcal{O}(n^{-1}), \qquad n \gg 1.
\end{equation}
The spacing is confirmed to be $\Delta\omega = -i/(2M)$ to all orders in the asymptotic expansion, validating Eq.~\eqref{spacing}.

This result connects to Hod's conjecture~\cite{Hod:1998vk}, which posits that the asymptotic QNM frequency of a BH encodes information about the BH area quantum. In its original formulation for Schwarzschild, the conjecture asserts $\omega_{\text{QNM}} \to \ln 3/(8\pi M) - i(n+1/2)/(4M)$, giving $|\Delta\omega_I| = 1/(4M)$. In the Kerr-EMDA case, the quasibound state spacing $|\Delta\omega_I| = 1/(2M) = 2\kappa_s^{\text{Sch}}$ (where $\kappa_s^{\text{Sch}} = 1/(4M)$ is the Schwarzschild surface gravity) provides a modification of the Hod conjecture that incorporates the dilaton and rotation.

The behavior of the real part also deserves attention. From Eq.~\eqref{asymptotic_freq}, $\omega_R^{(n)} \to -qQ/(2M) = -q\Phi_H^{\text{Sch}}$ as $n\to\infty$, where $\Phi_H^{\text{Sch}} = Q/(2M)$ is the electrostatic potential of the non-rotating dilaton BH (GMGHS limit). This means that highly damped modes ``forget'' about the BH spin and probe only the charge structure. The transition from spin-dependent to spin-independent behavior occurs at $n_{\text{crit}} \sim 2Ma\omega/\kappa_s$, beyond which the orbital contribution $\mathcal{L}$ becomes negligible.

In Table~\ref{tab:resonant}, we list the first several resonant frequencies obtained by numerically solving the implicit $\varepsilon_n$-condition~\eqref{epsiloncondition}. The spacing $|\Delta\omega_I|$ converges rapidly to $1/(2M) = 0.5$ with increasing $n$, while $\omega_R$ approaches the asymptotic value $-qQ/(2M) = -0.030$. The $\mathcal{O}(1/n)$ corrections are visible for low $n$ but become negligible by $n \gtrsim 5$.

\begin{table}[htbp]
\centering
\caption{Resonant frequencies $\omega_n$ from the $\varepsilon_n$-condition for $M=1$, $a=0.5M$, $Q=0.3M$, $q=0.2$, $\mu_s=0.5M^{-1}$, $m=1$, $\lambda_{\ell m}=2$. Values are obtained by numerically solving the implicit equation~\eqref{omega_implicit}. The last column shows $|\Delta\omega_I|$ converging to $1/(2M) = 0.5$, confirming Eq.~\eqref{spacing}. For comparison, the leading-order asymptotic formula~\eqref{asymptotic_freq} predicts $\omega_R = -0.030$ and $\omega_I^{(0)} = -1.773$ for all $n$.}
\label{tab:resonant}
\begin{tabular}{c c c c}
\hline\hline
$n$ & $\omega_R^{(n)}\,M$ & $\omega_I^{(n)}\,M$ & $|\Delta\omega_I|\,M$ \\
\hline
0 & $+0.0131$ & $-1.5388$ & --- \\
1 & $+0.0045$ & $-2.0248$ & 0.486 \\
2 & $-0.0016$ & $-2.5177$ & 0.493 \\
3 & $-0.0060$ & $-3.0136$ & 0.496 \\
4 & $-0.0092$ & $-3.5111$ & 0.497 \\
5 & $-0.0117$ & $-4.0094$ & 0.498 \\
\hline\hline
\end{tabular}
\end{table}

\section{Effective Potential and Scattering Structure} \label{sec5}

The radial equation~\eqref{radialODE} can be recast in a Schr\"{o}dinger-like form that makes the scattering physics transparent. In this section, we introduce the tortoise coordinate, define the effective potential, and analyze its structure as a function of the BH and field parameters.

\subsection{Tortoise coordinate and Schr\"{o}dinger-like form} \label{sec5a}

We define the tortoise coordinate $r_*$ through
\begin{equation}\label{tortoise}
\frac{dr_*}{dr} = \frac{r(r-2D)+a^2}{\Delta} = \frac{\Sigma_r}{\Delta}\,,
\end{equation}
where $\Sigma_r \equiv r(r-2D)+a^2$. This maps the outer horizon $r_+$ to $r_*\to -\infty$ and spatial infinity $r\to\infty$ to $r_*\to +\infty$. Setting $R(r) = \Psi_s(r)/\sqrt{\Sigma_r}$ and using Eq.~\eqref{tortoise}, the radial equation~\eqref{radialODE} transforms into
\begin{equation}\label{schrodinger}
\frac{d^2\Psi_s}{dr_*^2} + V_{\text{eff}}(r)\,\Psi_s = 0\,,
\end{equation}
where the effective potential is
\begin{equation}\label{Veff}
V_{\text{eff}}(r) = \frac{\Delta}{\Sigma_r^2}\!\left[\frac{\mathcal{K}^2}{\Delta} + 2am\omega - \lambda_{\ell m} - \mu_s^2\,r(r-2D) - a^2\omega^2\right] - \frac{\Delta}{\Sigma_r}\frac{d}{dr}\!\left(\frac{\Delta}{\Sigma_r}\frac{d\sqrt{\Sigma_r}}{dr}\right)\frac{1}{\sqrt{\Sigma_r}}\,.
\end{equation}
The first group of terms is the ``physical'' potential arising from the wave dynamics; the second is a curvature correction from the coordinate transformation. For the purpose of understanding the scattering structure, the dominant contribution comes from the first group. In the $s$-wave ($\ell = m = 0$) massless ($\mu_s = 0$) uncharged ($q = 0$) limit, the effective potential simplifies considerably and is strictly positive outside the horizon, forming a single barrier that governs the transmission of Hawking quanta.

The boundary conditions in the tortoise coordinate take the transparent form:
\begin{equation}\label{bcschrodinger}
\Psi_s \sim \begin{cases} e^{-i(\omega - m\Omega_H + q\Phi_H)\,r_*} & r_*\to -\infty \;\;(\text{ingoing at horizon}),\\[4pt]
\mathcal{R}\,e^{-i\tilde{k}r_*} + \mathcal{T}\,e^{+i\tilde{k}r_*} & r_*\to +\infty\;\;(\text{outgoing + reflected at infinity}),
\end{cases}
\end{equation}
where $\tilde{k} = \sqrt{\omega^2 - \mu_s^2}$ for scattering states ($\omega > \mu_s$), and $\mathcal{R}$, $\mathcal{T}$ are the reflection and transmission amplitudes.

\subsection{Effective potential analysis} \label{sec5b}

To isolate the main features of $V_{\text{eff}}$, we examine the dominant contribution (dropping the curvature correction, which is suppressed at large $r$). The asymptotic limits are:

\medskip
\noindent(i) \emph{At the horizon} ($r\to r_+$, $\Delta\to 0$): $V_{\text{eff}} \to (\omega - m\Omega_H + q\Phi_H)^2 \geq 0$, which vanishes at the superradiant threshold.

\medskip
\noindent(ii) \emph{At spatial infinity} ($r\to\infty$): $V_{\text{eff}} \to \omega^2 - \mu_s^2$. For $\omega > \mu_s$ (scattering states), $V_{\text{eff}}\to \tilde{k}^2 > 0$; for $\omega < \mu_s$ (bound states), $V_{\text{eff}}\to -|\varpi|^2 < 0$.

Between these limits, $V_{\text{eff}}$ generically develops a positive barrier whose height and width depend on the angular momentum $\ell$, spin $a$, dilaton parameter $D$, and the field charges $(q, \mu_s)$. For $\ell\geq 1$, the centrifugal barrier $\sim \ell(\ell+1)/r^2$ dominates at intermediate radii, producing a tall peak. The dilaton parameter $D$ lowers the peak height relative to the Kerr value: since $\Delta_{\text{Kerr-EMDA}}$ has roots shifted outward by $D$, the barrier is broader but shallower. This effect enhances the transmission of low-energy modes and leads to a larger GF, as we confirm quantitatively in Sec.~\ref{sec6}.

For massive fields ($\mu_s > 0$), an additional potential well develops outside the barrier when $\omega < \mu_s$, creating a trapping region that supports the quasibound states studied in Sec.~\ref{sec4}. The depth of this well increases with $\mu_s$, and the resonant frequencies $\omega_n$ correspond to quasi-stationary states tunneling through the barrier.

The scalar charge $q$ enters $V_{\text{eff}}$ through $\mathcal{K}(r)$, effectively shifting the frequency seen at the horizon to $\omega - q\Phi_H$. For $q\Phi_H > 0$ (same-sign charges), this reduces the effective frequency at the horizon and consequently lowers the barrier on the horizon side. This asymmetric deformation of the potential enhances absorption for co-charged scalars and suppresses it for counter-charged ones. These features are illustrated in Fig.~\ref{fig:Veff}: panel~(a) shows the $\ell$-dependence of the centrifugal barrier, while panel~(b) demonstrates the dilaton-induced lowering and broadening explicitly.

\subsection{Unstable null orbits and the potential peak} \label{sec5c}

In the geometric optics limit ($\omega \gg \mu_s$, $\ell \gg 1$), the peak of $V_{\text{eff}}$ corresponds to the location of unstable null circular orbits (the ``photon sphere'' in the non-rotating limit). For the Kerr-EMDA metric, the radius of the photon sphere in the equatorial plane ($\theta = \pi/2$) for co-rotating orbits satisfies~\cite{Ghosh:2020spb}
\begin{equation}\label{photonsphere}
r_{\text{ph}}^2 - 3Mr_{\text{ph}} + 2Dr_{\text{ph}} + 2a\sqrt{Mr_{\text{ph}} - D r_{\text{ph}}} = 0\,.
\end{equation}
In the Schwarzschild limit ($a=D=0$), Eq.~\eqref{photonsphere} gives $r_{\text{ph}} = 3M$. The dilaton parameter shifts the photon sphere inward: for $D > 0$, the coefficient of $r_{\text{ph}}$ changes from $-3M$ to $-3M+2D$, so $r_{\text{ph}}$ decreases. This inward shift concentrates the barrier closer to the horizon, facilitating tunneling and increasing the GF.

The connection between the potential peak and the QNM spectrum is well established~\cite{Cardoso:2008bp}: in the eikonal limit, the real part of the QNM frequency is related to the orbital frequency at $r_{\text{ph}}$, while the imaginary part is governed by the Lyapunov exponent of the null orbit. The dilaton-induced modifications to $r_{\text{ph}}$ therefore propagate directly into the QNM spectrum, providing an observable signature of the dilaton coupling.

\begin{figure}[htbp]
\centering
\begin{subfigure}[t]{0.48\textwidth}
\centering
\includegraphics[width=\textwidth]{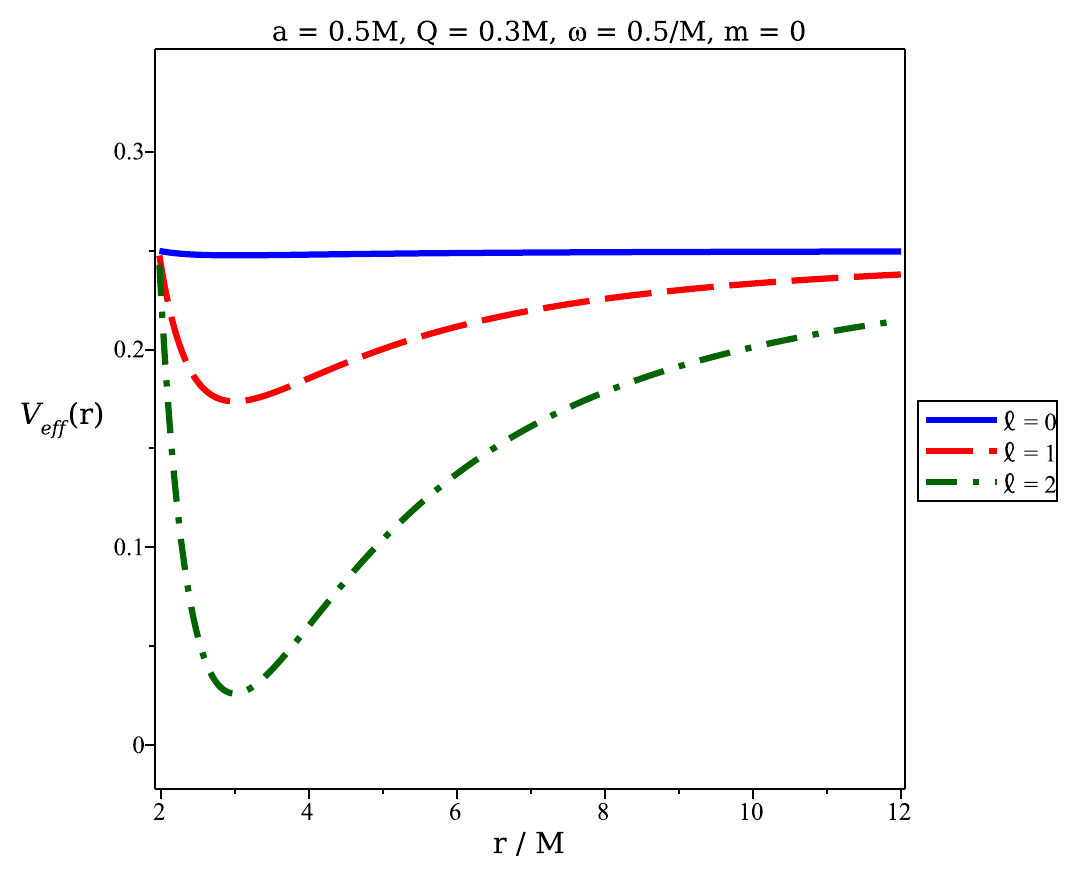}
\caption{Angular momentum dependence.}
\label{fig:Veff_ell}
\end{subfigure}
\hfill
\begin{subfigure}[t]{0.48\textwidth}
\centering
\includegraphics[width=\textwidth]{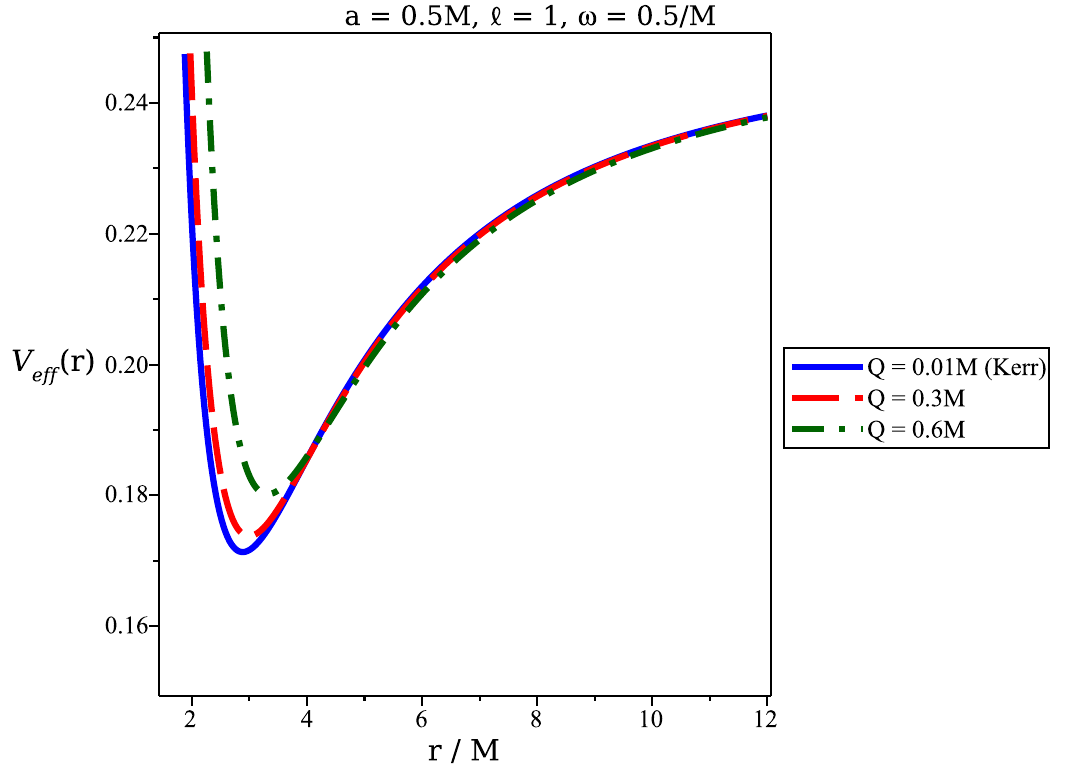}
\caption{Dilaton effect.}
\label{fig:Veff_dilaton}
\end{subfigure}
\caption{Effective potential $V_{\text{eff}}(r)$ for massless uncharged scalars on the Kerr-EMDA background with $M=1$ and $\omega = 0.5\,M^{-1}$. (a) $V_{\text{eff}}$ for $\ell = 0, 1, 2$ at fixed $a = 0.5M$, $Q = 0.3M$, $m = 0$: higher $\ell$ produces a deeper well and taller centrifugal barrier. (b) $V_{\text{eff}}$ for $Q = 0.01M$ (near-Kerr), $0.3M$, $0.6M$ at fixed $a = 0.5M$, $\ell = 1$: increasing $Q$ (and hence the dilaton parameter $D$) lowers the barrier minimum and shifts the peak outward, rendering the BH more transparent to scalar radiation.}
\label{fig:Veff}
\end{figure}

\section{Greybody Factor} \label{sec6}

{\color{black}In this section, we derive analytical GFs for the Kerr-EMDA BH.
We begin with the massless uncharged limit ($\mu_s = 0$, $q = 0$), where the CHF
reduces to the Gauss hypergeometric function ${}_2F_1$, and subsequently extend
the result to massless charged scalars ($\mu_s = 0$, $q \neq 0$) in Sec.~\ref{sec6e},
where we show that the same reduction applies for arbitrary $q$.}

\subsection{HeunC $\to$ ${}_2F_1$ reduction} \label{sec6a}

{\color{black}In the massless uncharged limit ($q=0$, $\mu_s = 0$), the Heun
parameters~\eqref{alphatilde}--\eqref{etatilde} simplify dramatically.
Setting $\mu_s = 0$ gives $\varpi^2 = -\omega^2$, so the square root
becomes $\sqrt{\mu_s^2 - \omega^2} = \pm i\omega$ and the irregular
singularity parameter reduces to $\tilde{\alpha} = \mp 2\omega\delta_r$,}
which is real and finite. Meanwhile, the function $\mathcal{K}(r)$
with $q = 0$ takes the form
\begin{equation}\label{K_neutral}
\mathcal{K}(r)\big|_{q=0} = \omega\,\Sigma_r - am
= \omega[r(r-2D) + a^2] - am\,,
\end{equation}
and the Frobenius exponents become
\begin{equation}\label{betagamma_neutral}
\tilde{\beta}\big|_{q=0}
= \frac{2i(2M\omega\,r_- - am)}{\delta_r}\,, \qquad
\tilde{\gamma}\big|_{q=0}
= -\frac{2i(2M\omega\,r_+ - am)}{\delta_r}.
\end{equation}

To make the reduction explicit, we change variables from $\zeta$ to $z = 1 - \zeta = (r_+ - r)/\delta_r$. With $\mu_s = q = 0$, the radial equation reduces to the hypergeometric equation
\begin{equation}\label{hypergeometric}
z(1-z)\frac{d^2F}{dz^2} + [c_0 - (a_0+b_0+1)z]\frac{dF}{dz} - a_0\,b_0\,F = 0\,,
\end{equation}
where the hypergeometric parameters are (see Appendix~\ref{appB} for details)
\begin{align}
a_0 &= \tfrac{1}{2}(1+\tilde{\gamma}-\tilde{\beta}) + i\omega\delta_r\,,\label{a0}\\
b_0 &= \tfrac{1}{2}(1+\tilde{\gamma}-\tilde{\beta}) - i\omega\delta_r\,,\label{b0}\\
c_0 &= 1+\tilde{\gamma}\,.\label{c0}
\end{align}
The solution regular at $z=0$ (the outer horizon) is
\begin{equation}\label{F2F1}
F(z) = {}_2F_1(a_0, b_0;\, c_0;\, z)\,.
\end{equation}

\subsection{Analytical greybody factor} \label{sec6b}

The GF $\Gamma_\ell(\omega)$ is the transmission probability: $\Gamma_\ell = |\mathcal{T}|^2$, where $\mathcal{T}$ is the transmission amplitude. To extract it, we connect the near-horizon solution~\eqref{F2F1} to the asymptotic ($r\to\infty$, i.e.\ $z\to -\infty$) form using the standard connection formula for ${}_2F_1$~\cite{DLMF}:
\begin{equation}\label{connection}
{}_2F_1(a_0,b_0;c_0;z) = \frac{\Gamma(c_0)\Gamma(b_0-a_0)}{\Gamma(b_0)\Gamma(c_0-a_0)}(-z)^{-a_0} + \frac{\Gamma(c_0)\Gamma(a_0-b_0)}{\Gamma(a_0)\Gamma(c_0-b_0)}(-z)^{-b_0}\,.
\end{equation}
As $r\to\infty$ ($z\to -\infty$), the two terms correspond to the outgoing and ingoing waves, respectively. After extracting the oscillatory phases and using $|\mathcal{R}|^2 + |\mathcal{T}|^2 = 1$ (valid for $\omega > m\Omega_H$), the GF takes the closed form~\cite{Harmark:2007jy,Sakalli:2022swm,Creek:2006ia}
\begin{equation}\label{GF}
{\Gamma_\ell(\omega) = 1 - \left|\frac{\Gamma(a_0)\,\Gamma(b_0)\,\Gamma(c_0-a_0-b_0)}{\Gamma(c_0-a_0)\,\Gamma(c_0-b_0)\,\Gamma(a_0+b_0-c_0)}\right|^2\,.}
\end{equation}
Since $c_0 - a_0 - b_0 = \tilde{\beta}$, this can also be written as
\begin{equation}\label{GF2}
\Gamma_\ell(\omega) = 1 - \left|\frac{\Gamma(a_0)\,\Gamma(b_0)\,\Gamma(\tilde{\beta})}{\Gamma(c_0-a_0)\,\Gamma(c_0-b_0)\,\Gamma(-\tilde{\beta})}\right|^2\,.
\end{equation}
This is the first analytical GF formula for the Kerr-EMDA BH.

\subsection{Limiting behaviors} \label{sec6c}

\noindent\textbf{Low-energy limit} ($\omega r_+ \ll 1$). In this regime, the hypergeometric parameters satisfy $|a_0|, |b_0| \ll 1$ for the lowest partial waves, and the GF admits the expansion~\cite{Das:1996we}
\begin{equation}\label{lowenergy}
\Gamma_\ell(\omega) \sim (\omega r_+)^{2\ell+2}\,\mathcal{C}_\ell(a,D)\,,\qquad \omega r_+\ll 1\,,
\end{equation}
where $\mathcal{C}_\ell$ is a coefficient depending on the BH parameters. For $\ell = 0$ ($s$-wave), the leading term is $\Gamma_0 \propto (\omega r_+)^2$. This power-law suppression reflects the increasing difficulty of tunneling through the centrifugal barrier as $\ell$ increases. The $s$-wave contribution dominates at low energies, consistent with the universal absorption result discussed in Sec.~\ref{sec7b}.

\medskip
\noindent\textbf{High-energy limit} ($\omega r_+ \gg 1$). The effective potential becomes negligible compared to $\omega^2$, and all partial waves are fully transmitted: $\Gamma_\ell \to 1$. The GF approaches unity from below, with the approach rate governed by the height of the potential barrier.

\medskip
\noindent\textbf{$s$-wave at threshold.} For $\omega \to m\Omega_H$ (the superradiant threshold with $q=0$), $\tilde{\gamma} \to 0$, so $c_0 \to 1$ and the Gamma function ratios simplify. In this limit, $\Gamma_0 \to 4\pi\omega T_H + \mathcal{O}(\omega^2)$, which is the signature of the ``area theorem'' for absorption: the low-energy absorption cross-section approaches the horizon area.

\subsection{Parametric analysis} \label{sec6d}

The analytical GF~\eqref{GF} allows a systematic study of how the Kerr-EMDA parameters modify the transmission probability relative to the Kerr case.

\medskip
\noindent\textbf{Dilaton effect.} The dilaton parameter $D = Q^2/(2M)$ enters the GF through three channels: (i) the shifted horizon radii $r_\pm$, which modify $\delta_r$ and hence $\tilde{\beta}$, $\tilde{\gamma}$; (ii) the modified $\Sigma_+$, which changes $\Omega_H$; and (iii) the altered surface gravity $\kappa_s$. For fixed $(M,a)$, increasing $Q$ (and hence $D$) shifts the outer horizon outward and increases the gap $\delta_r$, which broadens the effective potential barrier but lowers its peak. The net effect is an \emph{enhancement} of the GF at low energies: the Kerr-EMDA BH is more transparent to low-frequency scalars than the corresponding Kerr BH. This enhancement is largest for modes with $\ell = 0$ and diminishes with increasing $\ell$, since the centrifugal barrier dominates over the dilaton correction for high angular momentum. The dilaton effect on the $s$-wave GF is shown in Fig.~\ref{fig:GF_dilaton}: the full-range plot (a) confirms that all three values of $Q$ yield $\Gamma_0 \to 1$ at high frequencies, while the zoomed panel (b) reveals that larger $D$ produces a systematically higher $\Gamma_0$ in the low-energy tail, consistent with the barrier lowering seen in Fig.~\ref{fig:Veff}(b).

\begin{figure}[htbp]
\centering
\begin{subfigure}[t]{0.48\textwidth}
\centering
\includegraphics[width=\textwidth]{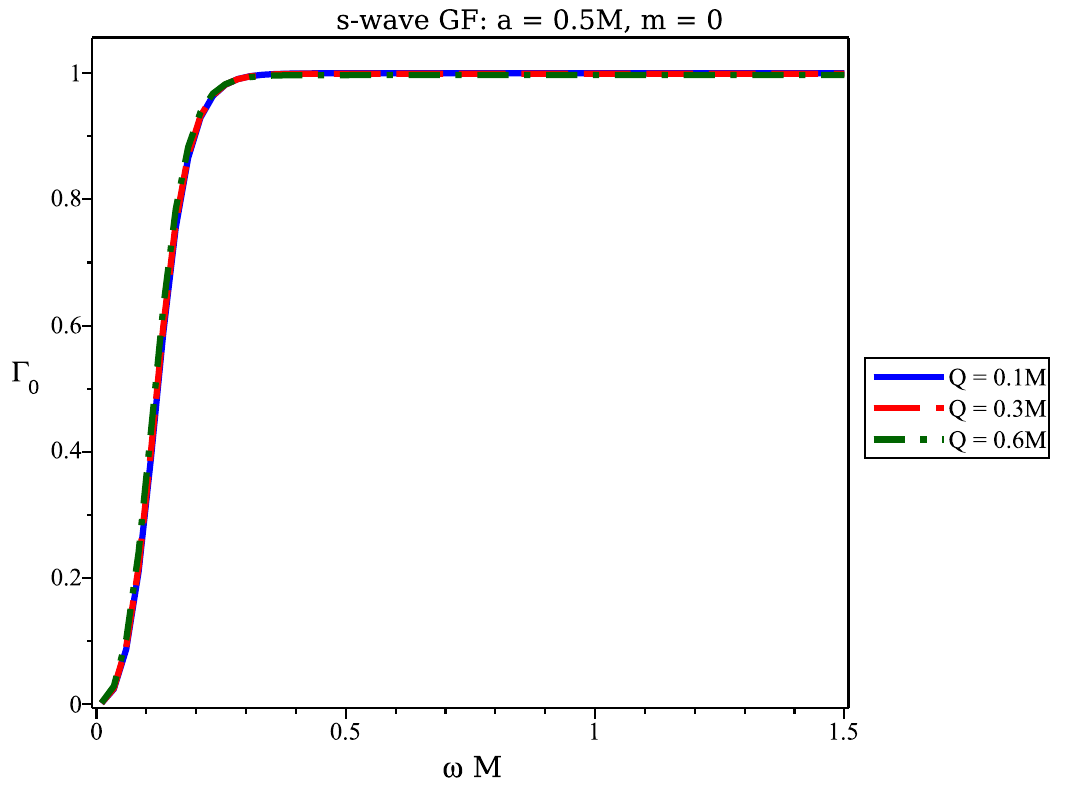}
\caption{Full range.}
\label{fig:GF_dilaton_full}
\end{subfigure}
\hfill
\begin{subfigure}[t]{0.48\textwidth}
\centering
\includegraphics[width=\textwidth]{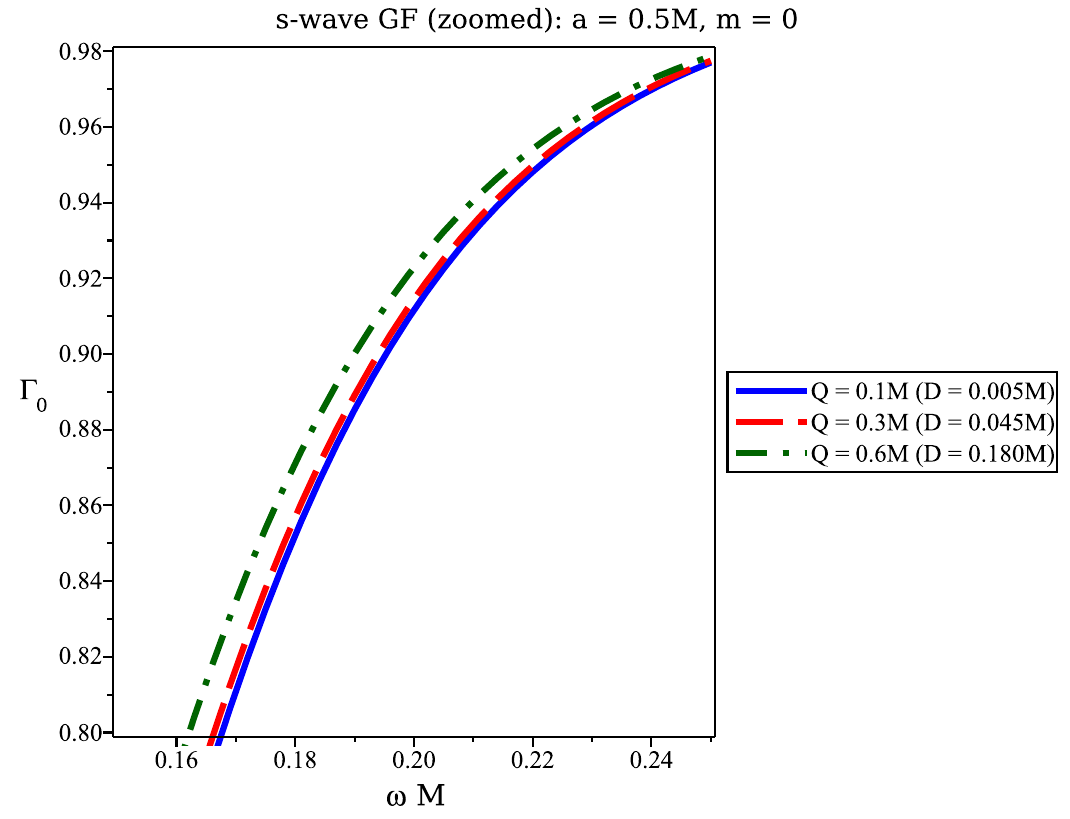}
\caption{Zoomed low-frequency region.}
\label{fig:GF_dilaton_zoom}
\end{subfigure}
\caption{$s$-wave ($\ell = 0$, $m = 0$) greybody factor $\Gamma_0(\omega)$ for the Kerr-EMDA BH with $a = 0.5M$, showing the dilaton effect for $Q = 0.1M$ ($D = 0.005M$), $0.3M$ ($D = 0.045M$), and $0.6M$ ($D = 0.180M$). (a) Full range: all curves saturate to unity at high frequencies. (b) Zoomed to $\omega M \in [0.15, 0.25]$: increasing $D$ enhances $\Gamma_0$ at low energies, confirming that the dilaton-broadened barrier is more transparent.}
\label{fig:GF_dilaton}
\end{figure}

\medskip
\noindent\textbf{Spin effect.} Increasing $a$ at fixed $(M,Q)$ has two competing effects: it reduces $r_+$ (compactifying the BH) and increases $\Omega_H$ (enhancing the rotational frame-dragging). The former tends to increase $\Gamma_\ell$ by thinning the barrier, while the latter introduces superradiant enhancement for $\omega < m\Omega_H$, where $\Gamma_\ell > 1$. For co-rotating modes ($m > 0$), these effects compound to produce a significant increase in the GF, while counter-rotating modes ($m < 0$) experience a suppressed GF. Figure~\ref{fig:GF_spin} illustrates the spin dependence for $m=0$: the full-range plot (a) shows that higher spin produces a slightly steeper rise of $\Gamma_0$ toward unity, while the zoomed panel (b) clearly resolves the hierarchy $\Gamma_0(a=0.8M) > \Gamma_0(a=0.5M) > \Gamma_0(a=0.3M) > \Gamma_0(a=0.1M)$ at intermediate frequencies.

\begin{figure}[htbp]
\centering
\begin{subfigure}[t]{0.48\textwidth}
\centering
\includegraphics[width=\textwidth]{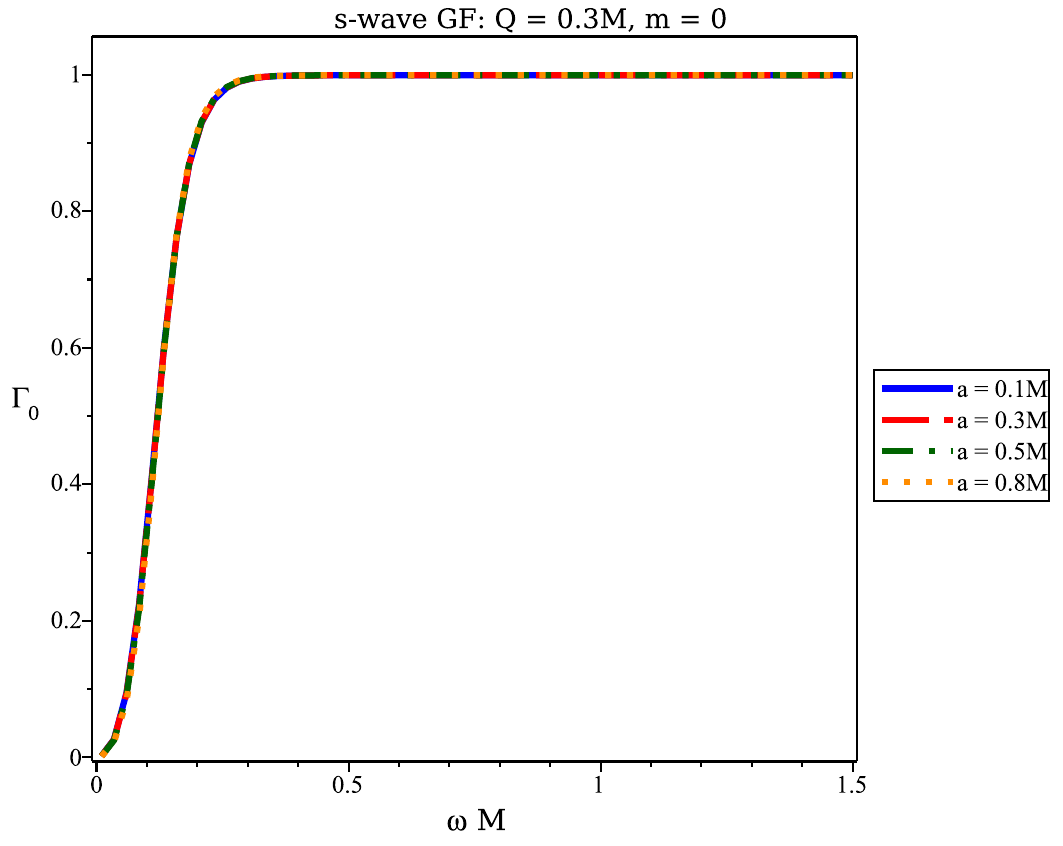}
\caption{Full range.}
\label{fig:GF_spin_full}
\end{subfigure}
\hfill
\begin{subfigure}[t]{0.48\textwidth}
\centering
\includegraphics[width=\textwidth]{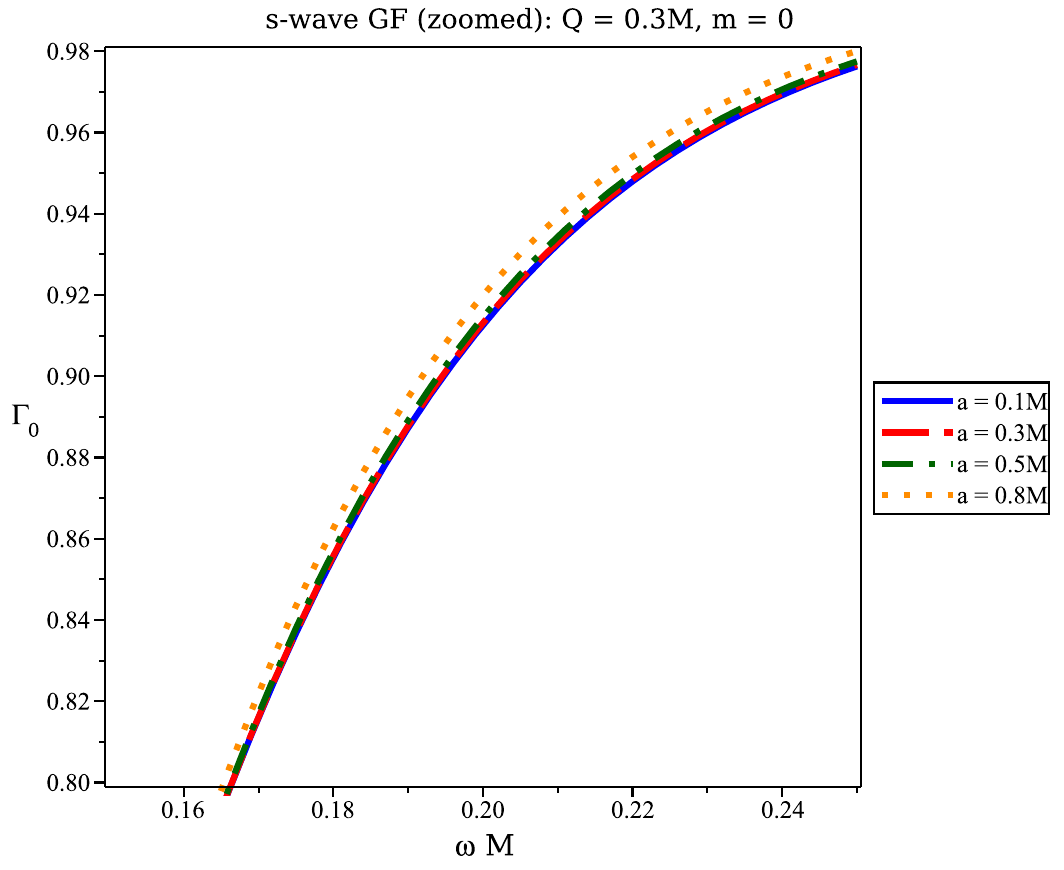}
\caption{Zoomed low-frequency region.}
\label{fig:GF_spin_zoom}
\end{subfigure}
\caption{$s$-wave ($\ell = 0$, $m = 0$) greybody factor $\Gamma_0(\omega)$ for the Kerr-EMDA BH with $Q = 0.3M$, showing the spin effect for $a = 0.1M$, $0.3M$, $0.5M$, and $0.8M$. (a) Full range: all curves approach unity at high frequencies. (b) Zoomed to $\omega M \in [0.15, 0.25]$: higher spin produces a larger GF, consistent with a thinner effective potential barrier.}
\label{fig:GF_spin}
\end{figure}

\medskip
\noindent\textbf{Partial wave hierarchy.} The GF is strongly $\ell$-dependent: $\Gamma_0 \gg \Gamma_1 \gg \Gamma_2$ at low energies, with the hierarchy relaxing at high energies where all $\Gamma_\ell \to 1$. The transition energy scale is set by the angular momentum barrier height, which grows as $\ell(\ell+1)$. This behavior is displayed in Fig.~\ref{fig:GF_ell}, where the steep rise of $\Gamma_\ell$ shifts progressively to higher $\omega$ with increasing $\ell$, reflecting the growing opacity of the centrifugal barrier.

\begin{figure}[htbp]
\centering
\includegraphics[width=0.55\textwidth]{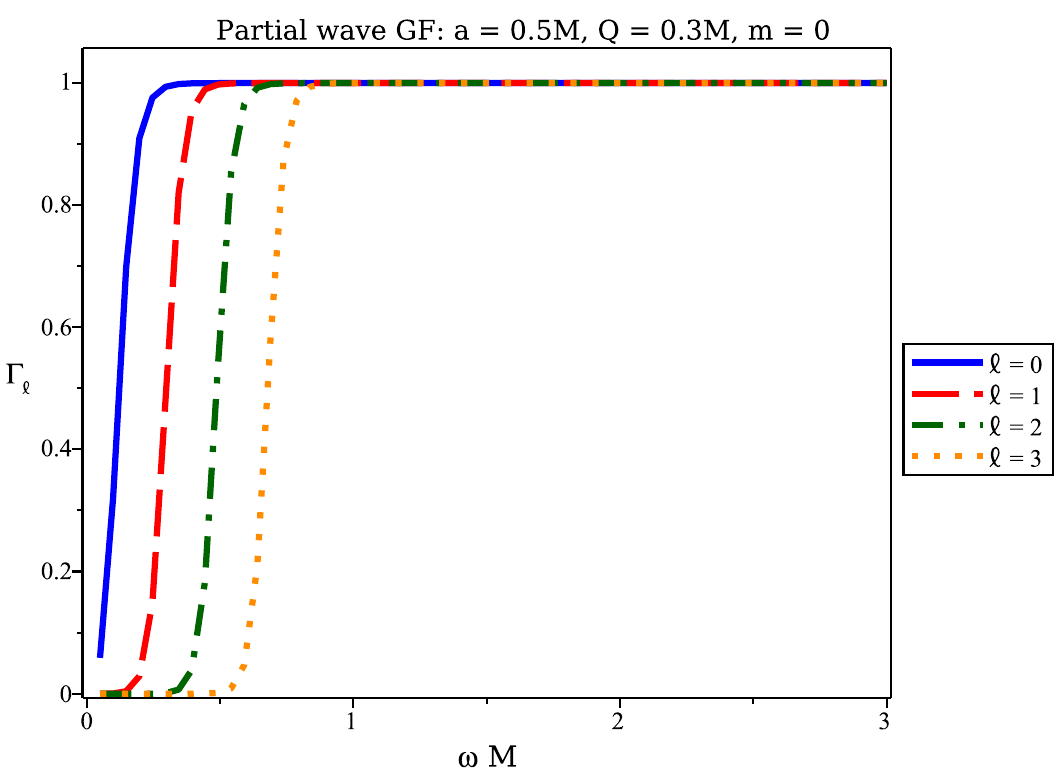}
\caption{Partial wave hierarchy of the greybody factor $\Gamma_\ell(\omega)$ for $\ell = 0, 1, 2, 3$ with $a = 0.5M$, $Q = 0.3M$, and $m = 0$. At low energies, $\Gamma_0 \gg \Gamma_1 \gg \Gamma_2 \gg \Gamma_3$, confirming the dominance of the $s$-wave. All partial waves converge to $\Gamma_\ell \to 1$ at high frequencies, but the onset of full transmission is delayed to progressively higher $\omega$ with increasing $\ell$.}
\label{fig:GF_ell}
\end{figure}

\subsection{Extension to massless charged scalars} \label{sec6e}

{\color{black}The HeunC~$\to$~${}_2F_1$ reduction \cite{Sakalli:2016mnk} performed in
Sec.~\ref{sec6a} was presented for $\mu_s = q = 0$, but a closer
inspection reveals that only the condition $\mu_s = 0$ is required.
The parameters $\tilde{\alpha}$~\eqref{alphatilde} and
$\tilde{\delta}$~\eqref{deltatilde} are both independent of the scalar
charge $q$: in $\tilde{\alpha}$, the quantity
$\sqrt{\mu_s^2 - \omega^2}$ carries no $q$-dependence, and in
$\tilde{\delta}$, the terms involving $2am\omega - \lambda_{\ell m}
- a^2\omega^2$ likewise contain no explicit $q$. The scalar charge
enters the CHE exclusively through the Frobenius exponents
$\tilde{\beta}$ and $\tilde{\gamma}$ [Eqs.~\eqref{betatilde}
and~\eqref{gammatilde}], which encode the near-horizon behavior via
$\mathcal{K}_\pm = (2M\omega + qQ)\,r_\pm - am$. Consequently,
the proportionality relation
$\tilde{\delta}|_{\mu_s=0} = \tilde{\alpha}\,[\ldots]$ in
Eq.~\eqref{delta_alpha_relation} and the gauge transformation that
converts the CHE into the Gauss hypergeometric equation hold for
arbitrary $q$ when $\mu_s = 0$.

The analytical GF for massless charged scalars on the Kerr-EMDA
BH is therefore
\begin{equation}\label{GFcharged}
\Gamma_\ell^{(q)}(\omega) = 1 -
\left|\frac{\Gamma(a_0)\,\Gamma(b_0)\,\Gamma(\tilde{\beta})}
{\Gamma(c_0 - a_0)\,\Gamma(c_0 - b_0)\,
\Gamma(-\tilde{\beta})}\right|^2,
\end{equation}
where the hypergeometric parameters retain the structural form of
Eqs.~\eqref{a0}--\eqref{c0},
\begin{equation}\label{abc_charged}
a_0 = \tfrac{1}{2}(1+\tilde{\gamma}-\tilde{\beta}) + i\omega\delta_r\,,
\qquad
b_0 = \tfrac{1}{2}(1+\tilde{\gamma}-\tilde{\beta}) - i\omega\delta_r\,,
\qquad
c_0 = 1 + \tilde{\gamma}\,,
\end{equation}
but now with the $q$-dependent Frobenius exponents
\begin{equation}\label{betagamma_charged}
\tilde{\beta} =
\frac{2i\bigl[(2M\omega + qQ)\,r_- - am\bigr]}{\delta_r}\,,
\qquad
\tilde{\gamma} =
\frac{-2i\bigl[(2M\omega + qQ)\,r_+ - am\bigr]}{\delta_r}
= \frac{-2i\Sigma_+(\omega - m\Omega_H + q\Phi_H)}{\delta_r}\,.
\end{equation}
Setting $q = 0$ in Eq.~\eqref{betagamma_charged} recovers the neutral
exponents~\eqref{betagamma_neutral}, and Eq.~\eqref{GFcharged} reduces
to the uncharged GF~\eqref{GF2}. The identity
$c_0 - a_0 - b_0 = \tilde{\beta}$ remains valid for all $q$.

The scalar charge modifies the GF through two channels. First, the
Frobenius index $\tilde{\gamma}$ at the outer horizon acquires the
shift $-2iqQr_+/\delta_r$ relative to the neutral case, which
alters the near-horizon phase and hence the transmission amplitude.
Second, the inner-horizon index $\tilde{\beta}$ shifts by
$2iqQr_-/\delta_r$, affecting the Gamma-function ratio that
controls the reflection coefficient. For co-charged scalars
($qQ > 0$), $|\text{Im}(\tilde{\gamma})|$ increases
[cf.\ Table~\ref{tab:Heun}], enhancing absorption at
frequencies above the superradiant bound; for counter-charged
scalars ($qQ < 0$), the opposite trend holds.}

\subsection{Superradiant amplification} \label{sec6f}

{\color{black}Equation~\eqref{GFcharged} was derived under the flux
conservation relation $|\mathcal{R}|^2 + |\mathcal{T}|^2 = 1$, which
holds for $\omega > \omega_c$. To extend the GF to the superradiant
regime, one must account for the reversal of the group velocity at the
horizon when $\omega < \omega_c = m\Omega_H - q\Phi_H$. The
flux-normalized absorption probability is~\cite{Brito:2015oca}
\begin{equation}\label{GFphysical}
\Gamma_\ell^{(\mathrm{abs})}(\omega)
= \frac{\omega - m\Omega_H + q\Phi_H}{\omega}\,
\Gamma_\ell^{(q)}(\omega)\,,
\end{equation}
where $\Gamma_\ell^{(q)}$ is the Gamma-function formula~\eqref{GFcharged}.
For $\omega > \omega_c$, both factors in Eq.~\eqref{GFphysical} are
positive and $\Gamma_\ell^{(\mathrm{abs})}$ reduces to the standard
absorption probability. For $\omega < \omega_c$, the prefactor
$(\omega - m\Omega_H + q\Phi_H)/\omega < 0$ while
$\Gamma_\ell^{(q)} > 0$, yielding
$\Gamma_\ell^{(\mathrm{abs})} < 0$: the reflected wave carries
more energy than the incident one. The superradiant amplification factor,
\begin{equation}\label{Zfactor}
Z_\ell^{(q)}(\omega)
\equiv -\Gamma_\ell^{(\mathrm{abs})}(\omega) > 0\,,
\qquad \omega < \omega_c\,,
\end{equation}
measures the fractional energy gain per scattering event, drawn from
the rotational and electromagnetic energy of the BH. 
\begin{table}[htbp]
\centering
\caption{{\color{black}Superradiant amplification for massless charged scalars with
$M=1$, $a=0.5M$, $Q=0.3M$, and $m=1$. The test frequency is
$\omega = \omega_c - 0.01\,M^{-1}$ (just below the superradiant bound).
$\Gamma_\ell^{(q)}$ is the Gamma-function formula~\eqref{GFcharged},
$\Gamma_\ell^{(\mathrm{abs})}$ is the flux-normalized absorption
probability~\eqref{GFphysical}, and $Z_\ell^{(q)} = -\Gamma_\ell^{(\mathrm{abs})}$
is the amplification factor~\eqref{Zfactor}. Negative $\Gamma_\ell^{(\mathrm{abs})}$
confirms superradiant energy extraction. For $q = 1.0$, the bound
$\omega_c < 0$ and superradiance is quenched for positive real frequencies.}}
\label{tab:superradiance}
\begin{tabular}{c c c c c c}
\hline\hline
$q$ & $\omega_c\,M$ & $\omega\,M$ & $\Gamma_\ell^{(q)}$ &
$\Gamma_\ell^{(\mathrm{abs})}$ & $Z_\ell^{(q)}$ \\
\hline
$-0.5$ & $0.2024$ & $0.1924$ & $0.0852$ & $-0.0044$ & $0.0044$ \\
$0.0$  & $0.1274$ & $0.1174$ & $0.1291$ & $-0.0110$ & $0.0110$ \\
$0.2$  & $0.0974$ & $0.0874$ & $0.1438$ & $-0.0165$ & $0.0165$ \\
$0.5$  & $0.0524$ & $0.0424$ & $0.1590$ & $-0.0375$ & $0.0375$ \\
$1.0$  & $-0.0226$ & --- & --- & \multicolumn{2}{c}{superradiance is quenched} \\
\hline\hline
\end{tabular}
\end{table}

Table~\ref{tab:superradiance} lists the amplification factor for several
values of the scalar charge at fixed $a = 0.5M$ and $Q = 0.3M$, with the test frequency
set to $\omega = \omega_c - 0.01\,M^{-1}$ in each case. Three trends are apparent.
First, all entries satisfy $\Gamma_\ell^{(\mathrm{abs})} < 0$, confirming that the
flux-normalized formula~\eqref{GFphysical} correctly captures superradiant amplification
whenever $\omega < \omega_c$. Second, the amplification factor $Z_\ell^{(q)}$ grows
monotonically from $0.44\%$ at $q = -0.5$ to $3.75\%$ at $q = 0.5$; this increase
arises because larger same-sign $q$ pushes $\omega_c$ closer to zero, so the fixed
offset $\omega = \omega_c - 0.01$ probes lower frequencies where the group-velocity
ratio $|(\omega - \omega_c)/\omega|$ is enhanced. Third, for $q = 1.0$ the superradiant
bound turns negative ($\omega_c = -0.023\,M^{-1}$), confirming that the electromagnetic
coupling has quenched superradiance entirely for positive real frequencies: the scalar
charge exceeds the critical value $q_{\mathrm{cr}} = m\Omega_H/\Phi_H \simeq 0.849$
identified in Sec.~\ref{sec3d}.

Two limiting
behaviors merit attention:

\medskip
\noindent(i) \emph{Same-sign charges} ($qQ > 0$). The chemical-potential
term $-q\Phi_H < 0$ lowers $\omega_c$ below the neutral value
$m\Omega_H$, narrowing the superradiant window. For sufficiently large
$q > q_{\mathrm{cr}} = m\Omega_H/\Phi_H$, the bound $\omega_c$ becomes
negative and superradiance is quenched entirely for positive real
frequencies [cf.\ Fig.~\ref{fig:K_superrad}(b) and
Table~\ref{tab:Heun}, $q = 1.0$ row].

\medskip
\noindent(ii) \emph{Opposite-sign charges} ($qQ < 0$). Now
$-q\Phi_H > 0$ raises $\omega_c$ above $m\Omega_H$, widening the
superradiant window and increasing the peak amplification.

\medskip
For massive scalars ($\mu_s \neq 0$), the irregular singularity at
$\zeta \to \infty$ is genuine and cannot be removed by a gauge
transformation. A closed-form GF analogous to Eq.~\eqref{GFcharged}
does not exist in this case: the transmission coefficient depends on
the connection coefficients of the CHF between the regular singular
point $\zeta = 1$ (outer horizon) and the irregular singular point
at infinity, which are expressible only as infinite continued fractions
via the three-term recurrence~\eqref{recurrence}. Their numerical
evaluation, or the equivalent Painlev\'{e}/isomonodromic deformation
approach~\cite{Bonelli:2021uvf}, constitutes a natural direction for
future work.}

\section{Hawking Radiation Spectrum and Absorption Cross-Section} \label{sec7}

The analytical GF derived in the previous section enables us to compute the two principal observables of BH radiation as seen by a distant observer: the greybody-weighted Hawking spectrum and the frequency-dependent absorption cross-section. The bare thermal spectrum emitted at the horizon is filtered by the effective potential barrier, so that the flux reaching infinity is suppressed at low frequencies (where the barrier is opaque) and approaches the Planckian form only at high frequencies (where $\Gamma_\ell \to 1$). For the Kerr-EMDA BH, two competing dilaton effects shape the observable spectrum: the elevated Hawking temperature $T_H^{\text{EMDA}} > T_H^{\text{Kerr}}$ pushes the thermal peak to higher energies, while the enhanced GF amplifies the low-frequency tail. Together, these produce a higher total luminosity and a characteristic spectral distortion relative to Kerr that could, in principle, be used to constrain the dilaton parameter $D$. We also verify the low-energy universality of the absorption cross-section, $\sigma_{\text{abs}} \to \mathcal{A}_H$, and examine the dilaton-dependent oscillatory structure at intermediate frequencies that encodes information about the photon sphere.

\subsection{Greybody-weighted thermal spectrum} \label{sec7a}

The Hawking emission rate for massless, uncharged scalars from a Kerr-EMDA BH is obtained by weighting the thermal distribution with the GF~\eqref{GF}:
\begin{equation}\label{particlerate}
\frac{d^2N}{dt\,d\omega} = \sum_{\ell=0}^{\infty}\sum_{m=-\ell}^{+\ell}\frac{\Gamma_\ell(\omega)}{2\pi}\,\frac{1}{\exp\!\left[\frac{\omega - m\Omega_H}{T_H}\right] - 1}\,.
\end{equation}
For the charged case ($q\neq 0$), {\color{black}the thermal factor includes the chemical potential term
$\omega - m\Omega_H + q\Phi_H$}, {\color{black}the analytical GF for the charged case is derived in
Sec.~\ref{sec6e}.} The energy emission rate is
\begin{equation}\label{powerrate}
\frac{d^2E}{dt\,d\omega} = \omega\,\frac{d^2N}{dt\,d\omega}\,.
\end{equation}

Several features of the Kerr-EMDA Hawking spectrum distinguish it from the Kerr case. First, the higher Hawking temperature ($T_H^{\text{EMDA}} > T_H^{\text{Kerr}}$ for the same $M$ and $a$) shifts the peak of the thermal distribution to higher frequencies. Since $T_H \propto 1/\Sigma_+$ and $\Sigma_+^{\text{EMDA}} < \Sigma_+^{\text{Kerr}}$ (due to the $-2D$ term), the Kerr-EMDA BH radiates at a higher characteristic energy. Second, the enhanced GF (larger $\Gamma_\ell$ at low energies, as discussed in Sec.~\ref{sec6d}) further amplifies the emission at the low-frequency tail. These two effects combine to produce a \emph{higher total luminosity} for the Kerr-EMDA BH: dilaton BHs evaporate faster than their Kerr counterparts.

The superradiant regime $\omega < m\Omega_H$ produces stimulated emission: the Bose-Einstein factor becomes negative, and $\Gamma_\ell > 1$, leading to $d^2N/dt\,d\omega < 0$ (emission exceeding the incident flux). The dilaton parameter modifies the superradiant bandwidth through its effect on $\Omega_H$.

These features are confirmed by the numerical results presented in Figs.~\ref{fig:Hawking_N} and~\ref{fig:Hawking_E}. The particle emission spectrum (Fig.~\ref{fig:Hawking_N}) shows that increasing $Q$ raises the peak amplitude and shifts it to slightly higher frequencies, directly reflecting the elevated Hawking temperature. The energy emission rate (Fig.~\ref{fig:Hawking_E}), summed over $\ell = 0, 1, 2$, exhibits the same trend with an additional $\omega$-weighting that further separates the curves at intermediate frequencies. The total luminosity (area under each curve) increases with $Q$, confirming that Kerr-EMDA BHs evaporate faster than their Kerr counterparts.

\begin{figure}[htbp]
\centering
\begin{subfigure}[t]{0.48\textwidth}
\centering
\includegraphics[width=\textwidth]{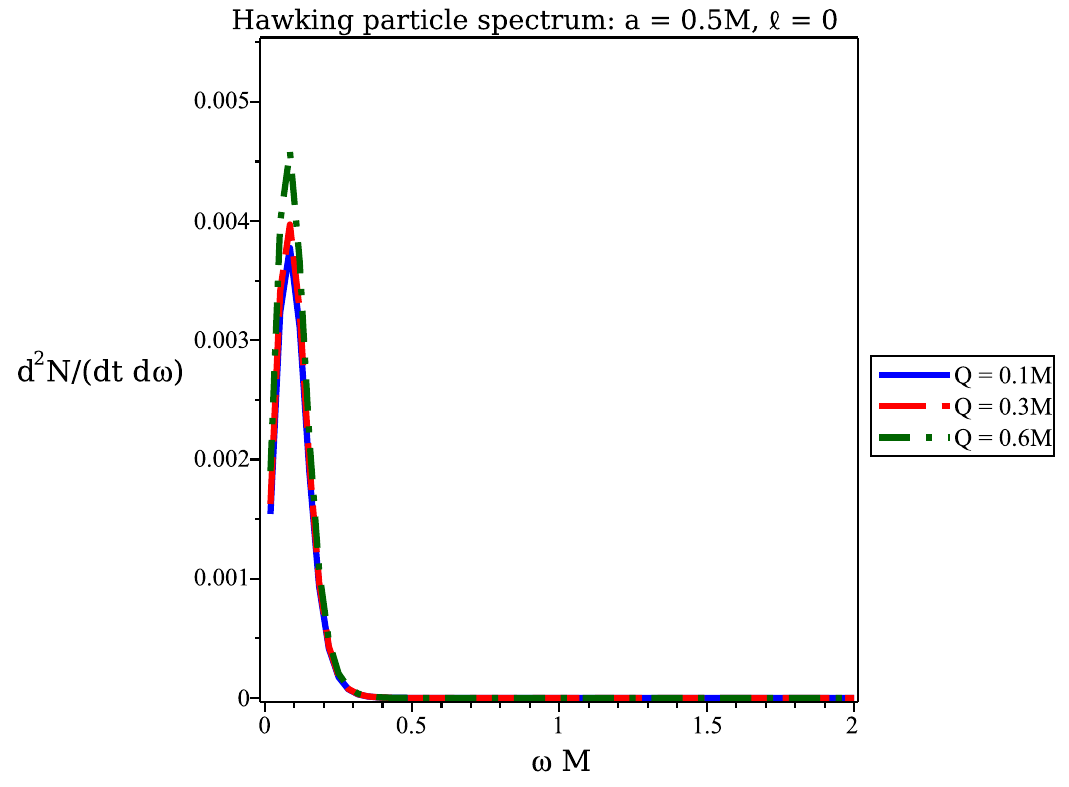}
\caption{Particle emission rate ($\ell = 0$).}
\label{fig:Hawking_N}
\end{subfigure}
\hfill
\begin{subfigure}[t]{0.48\textwidth}
\centering
\includegraphics[width=\textwidth]{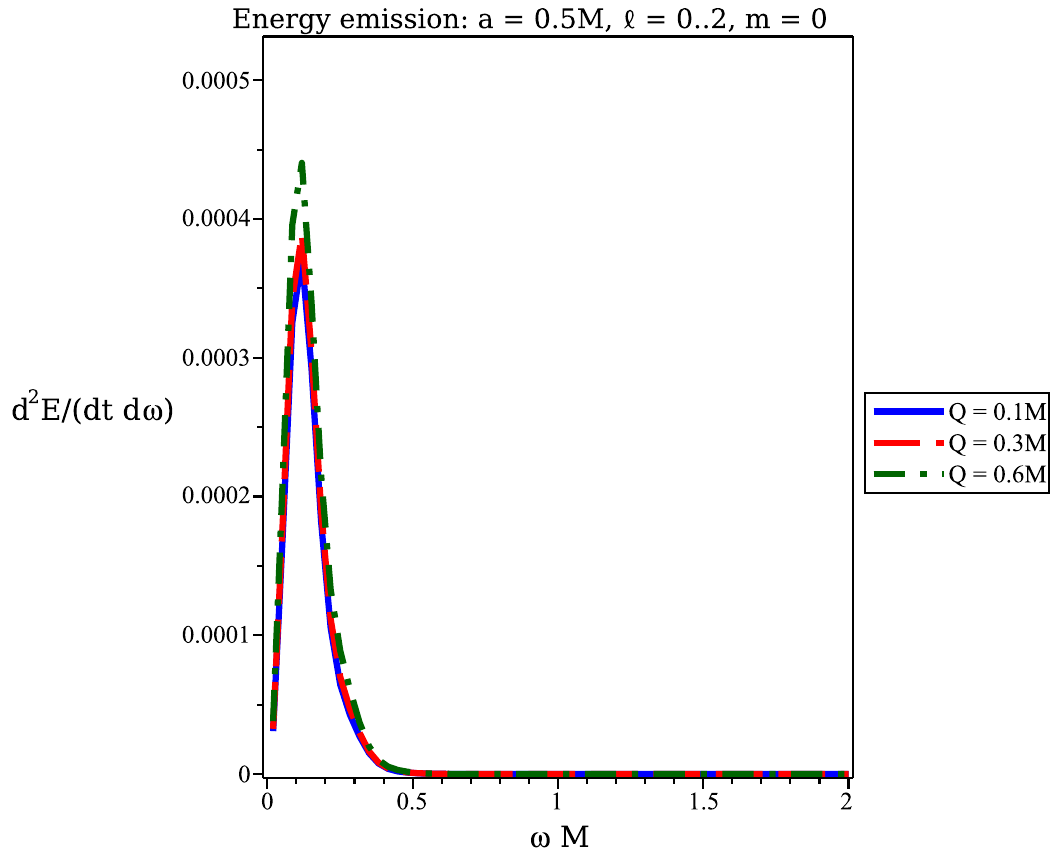}
\caption{Energy emission rate ($\ell = 0, 1, 2$).}
\label{fig:Hawking_E}
\end{subfigure}
\caption{Hawking radiation from the Kerr-EMDA BH with $a = 0.5M$ and $m = 0$, for $Q = 0.1M$, $0.3M$, and $0.6M$. (a) Particle emission rate $d^2N/(dt\,d\omega)$ for the $s$-wave: the peak shifts to higher $\omega$ and grows in amplitude with increasing $Q$, driven by the elevated Hawking temperature. (b) Energy emission rate $d^2E/(dt\,d\omega)$ summed over $\ell = 0, 1, 2$: the total luminosity increases with $Q$, confirming that dilaton-charged BHs evaporate faster.}
\label{fig:Hawking}
\end{figure}

\subsection{Absorption cross-section} \label{sec7b}

The partial-wave absorption cross-section for massless scalars is
\begin{equation}\label{abscs2}
\sigma_{\text{abs}}(\omega) = \sum_{\ell=0}^{\infty}\frac{\pi(2\ell+1)}{\omega^2}\,\Gamma_\ell(\omega)\,.
\end{equation}
In the low-energy limit, the $s$-wave dominates and the universality of BH absorption~\cite{Das:1996we,Higuchi:2001si} gives
\begin{equation}\label{universality}
\sigma_{\text{abs}}(\omega) \xrightarrow{\omega\to 0} \mathcal{A}_H = 4\pi\Sigma_+ = 4\pi\!\left[r_+(r_+-2D)+a^2\right]\,.
\end{equation}
This result follows from the fact that the $\ell = 0$ GF satisfies $\Gamma_0/\omega^2 \to \mathcal{A}_H/\pi$ as $\omega\to 0$, regardless of the specific BH geometry (provided it has a non-degenerate horizon). For the Kerr-EMDA BH, $\mathcal{A}_H$ is smaller than the Kerr value $4\pi(r_+^2+a^2)$ by the amount $8\pi D r_+$, reflecting the dilaton-induced reduction of the horizon area.

At intermediate energies, $\sigma_{\text{abs}}$ develops an oscillatory structure due to the interference between partial waves with different $\ell$. The oscillation period scales as $\Delta\omega \sim 1/r_{\text{ph}}$, where $r_{\text{ph}}$ is the photon sphere radius discussed in Sec.~\ref{sec5c}. The dilaton-induced inward shift of $r_{\text{ph}}$ increases the oscillation frequency, providing another observable handle to constrain $D$.

In the high-energy (geometric optics) limit, the absorption cross-section approaches the capture cross-section of the BH:
\begin{equation}\label{geometric}
\sigma_{\text{abs}}(\omega) \xrightarrow{\omega\to\infty} \sigma_{\text{geo}} = \pi b_c^2\,,
\end{equation}
where $b_c$ is the critical impact parameter determined by the unstable null orbit.

Figure~\ref{fig:sigma_abs} displays the normalized absorption cross-section $\sigma_{\text{abs}}/\mathcal{A}_H$ computed by summing the numerical GF over $\ell = 0, 1, 2, 3$. All three curves approach unity at low frequencies, verifying the universality relation~\eqref{universality}. At intermediate energies, the oscillatory structure from partial wave interference is clearly visible, with the oscillation amplitude and periodicity varying with $Q$. The overall envelope decreases as $\sim 1/\omega^2$ at high frequencies, as expected from Eq.~\eqref{abscs2} since $\Gamma_\ell \to 1$ while the prefactor decays as $\omega^{-2}$.

\begin{figure}[htbp]
\centering
\includegraphics[width=0.55\textwidth]{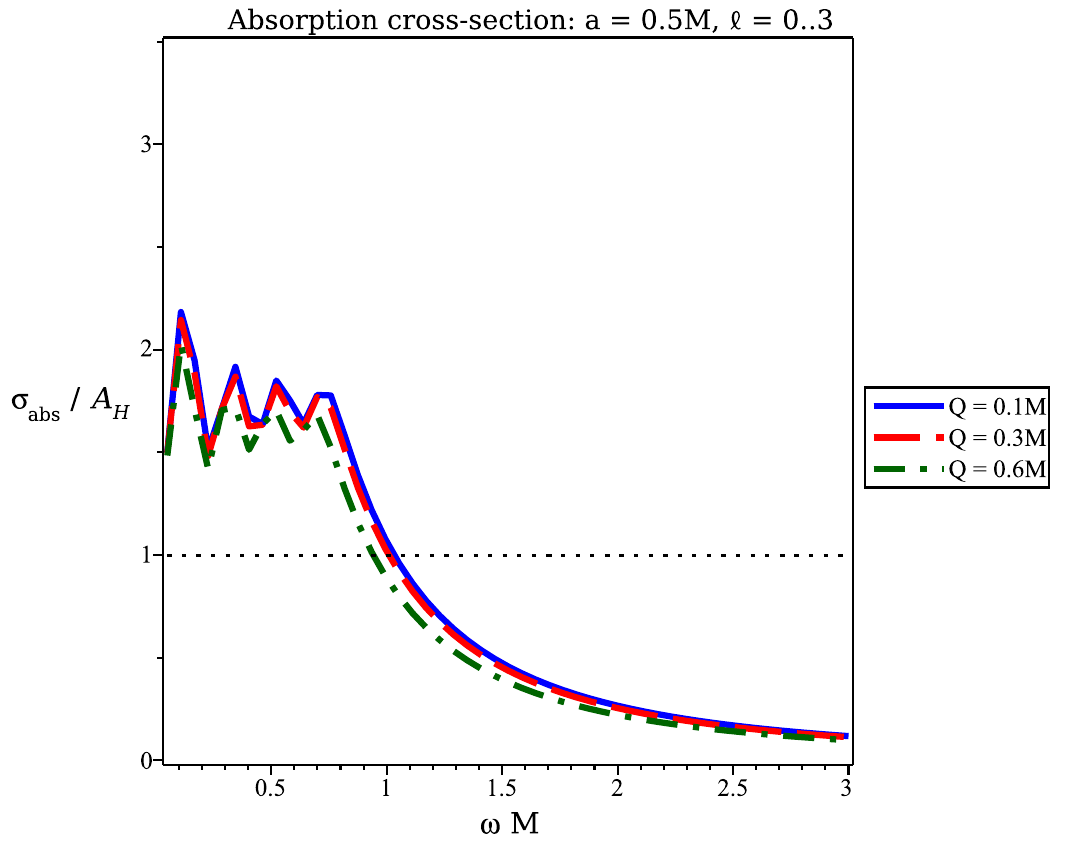}
\caption{Normalized absorption cross-section $\sigma_{\text{abs}}/\mathcal{A}_H$ versus $\omega M$, summed over $\ell = 0, 1, 2, 3$ with $a = 0.5M$ and $m = 0$, for $Q = 0.1M$, $0.3M$, and $0.6M$. The horizontal dotted line marks $\sigma_{\text{abs}} = \mathcal{A}_H$. All curves approach unity as $\omega \to 0$, verifying the low-energy universality. The oscillatory structure at intermediate energies reflects partial wave interference, with the pattern depending on the dilaton parameter $D$.}
\label{fig:sigma_abs}
\end{figure}

\section{Special Limits and Consistency Checks} \label{sec8}

The results of this paper encompass several well-known cases as special limits. In this section, we verify the consistency of our formalism by recovering established results in each limit.

\medskip
\noindent\textbf{(i) Kerr limit} ($D\to 0$, $Q\to 0$). Setting $D=Q=0$ eliminates the dilaton and Maxwell fields. The horizons reduce to $r_\pm = M \pm \sqrt{M^2-a^2}$, the function $\mathcal{K}$ becomes $\omega(r^2+a^2) - am$, and all Heun parameters reduce to the known Kerr values~\cite{Fiziev:2009,Vieira:2023EGB}. The GF~\eqref{GF} recovers the analytical Kerr GF~\cite{Harmark:2007jy}. The entropy quantum $\delta S_{\text{BH}} = 4\pi r_+/(r_+ - r_-)$ agrees with the Kerr result obtained via the Maggiore method~\cite{Maggiore:2007nq}.

\medskip
\noindent\textbf{(ii) Schwarzschild limit} ($a\to 0$, $D\to 0$). In this doubly degenerate limit, $r_+ = 2M$, $r_- = 0$, $\delta_r = 2M$, $\Sigma_+ = 4M^2$, and $\kappa_s = 1/(4M)$. The spacing becomes $|\Delta\omega_I| = 1/(2M) = 2\kappa_s$, and the entropy quantum gives $\delta S_{\text{BH}} = 4\pi$, corresponding to $\delta\mathcal{A}_H = 16\pi\ell_P^2$. This matches the classic Bekenstein-Maggiore result~\cite{Maggiore:2007nq}. The CHE reduces to the standard hypergeometric equation, and the GF reproduces the known Schwarzschild result~\cite{Sakalli:2022swm}.

\medskip
\noindent\textbf{(iii) Static EMDA (GMGHS) limit} ($a\to 0$). Setting $a=0$ with $Q\neq 0$ gives the static dilaton BH~\cite{Gibbons:1988rs,Garfinkle:1990qj}. The horizons become $r_+ = 2(M+D)$, $r_- = 0$, and $\Sigma_+ = r_+(r_+-2D) = 4M(M+D)$. The angular equation reduces to the associated Legendre equation. The radial equation retains its confluent Heun form for the massive case ($\mu_s \neq 0$), but simplifies to the hypergeometric equation for massless fields, consistent with the known GMGHS scattering analysis~\cite{Fernando:2005iz}.

\medskip
\noindent\textbf{(iv) Neutral scalar limit} ($q\to 0$). Setting $q=0$ recovers the setup of Senjaya and Ponglertsakul~\cite{Senjaya:2025kerremda}. The function $\mathcal{K}$ loses the $qQr$ term, and the Heun parameters simplify accordingly. Our Table~\ref{tab:Heun} rows with $q=0$ (Kerr and Neutral) serve as direct verification. The resonant frequency spacing $|\Delta\omega_I| = 1/(2M)$ remains unchanged, confirming that this is a property of the background geometry rather than the probe field.

\medskip
\noindent\textbf{(v) RLDBH comparison.} In the RLDBH framework of Ref.~\cite{Sakalli:2016mnk}, the resonant frequency spectrum has the form $\omega_n = m\Omega_H + q\Phi_e + i(n+1)\kappa_s$, yielding $|\Delta\omega_I| = \kappa_s$ and consequently $\delta S_{\text{BH}} = 2\pi$ universally. The key structural difference is that the RLDBH has $r_- = 0$ (effectively a single-horizon geometry), so $\delta_r = r_+$ and $r_+/\delta_r = 1$. In the Kerr-EMDA case, $r_+/\delta_r > 1$ generically (both horizons are non-zero), leading to $\delta S_{\text{BH}} > 2\pi$. The two results agree only in the Schwarzschild limit, where $r_+/\delta_r = 1$ and $\delta S_{\text{BH}} = 4\pi = 2\times 2\pi$.

\medskip
\noindent\textbf{(vi) Kerr-Newman comparison.} The Kerr-Newman BH in Einstein-Maxwell theory (without the dilaton) has $D = 0$ and $r_\pm = M \pm \sqrt{M^2 - a^2 - Q^2}$. The dilaton in EMDA theory replaces $Q^2$ by $2MD$ in the horizon structure, so the Kerr-EMDA horizons are \emph{not} obtained by simply substituting $Q^2 \to 2MD$ in the Kerr-Newman formula. Rather, the dilaton shift enters additively: $r_\pm = (M+D) \pm \sqrt{(M+D)^2 - a^2}$, which is equivalent to replacing $M \to M+D$ in the Kerr formula. This distinction has observable consequences: for the same $(M, a, Q)$, the Kerr-EMDA BH has larger $r_+$, smaller $\Sigma_+$, higher $T_H$, and a larger GF than the Kerr-Newman BH. {\color{black}Numerical QNM frequencies and GFs for massive charged
scalars on the Kerr-Newman background are available in
Refs.~\cite{Konoplya:2013rxa,Konoplya:2014sna,Konoplya:2007zx};
a direct comparison with the present Kerr-EMDA analytical results
would isolate the dilaton contribution and is left for future work.}

\section{Conclusions} \label{sec9}

In this work, we have carried out an analytical study of charged massive scalar field perturbations on the Kerr-EMDA BH background, extending the neutral-scalar treatment of Senjaya and Ponglertsakul~\cite{Senjaya:2025kerremda} to the physically richer case of electromagnetic coupling between the probe field and the background gauge potential. By solving the gauge-covariant KGE via separation of variables, we obtained the first exact solution for this system: both the radial and angular sectors are expressed in terms of CHFs, with all five Heun parameters given explicitly as functions of the BH parameters $(M, a, Q)$ and the field parameters $(\omega, \mu_s, q, m, \ell)$. The scalar charge $q$ modifies these parameters in a manner that cannot be absorbed by a simple redefinition of existing quantities, reflecting the genuinely new physics introduced by the electromagnetic coupling.

The CHF polynomial truncation condition yields a discrete tower of complex resonant frequencies whose imaginary parts {\color{black}are asymptotically equispaced} with $|\Delta\omega_I| = 1/(2M)$ {\color{black}in the highly-damped regime ($n \gg 1$)}. This spacing depends only on the BH mass and is independent of the spin $a$, charge $Q$, scalar charge $q$, field mass $\mu_s$, and angular momentum quantum numbers --- a universality that stems from the explicit $2M\omega$ coefficient in the $(\tilde{\beta}+\tilde{\gamma})/2$ term of the $\varepsilon_n$-condition{\color{black}; at finite $n$, a residual correction from the orbital contribution $\mathcal{L}$ is present but converges rapidly, reaching 99.6\% of the asymptotic value by $n = 5$ (Table~\ref{tab:resonant})}. In the highly-damped regime, the real part of the frequency converges to $\omega_R \to -qQ/(2M)$, indicating that highly damped modes probe only the electrostatic structure of the BH and become insensitive to its rotation. Through the Maggiore prescription and the first law of BH thermodynamics, the {\color{black}asymptotically} equispaced frequency spectrum translates into a quantized entropy spectrum $\delta S_{\text{BH}} = 4\pi r_+/(r_+ - r_-)$, which is parameter-dependent and diverges at extremality. This stands in contrast to the universal $\delta S_{\text{BH}} = 2\pi$ obtained for the RLDBH~\cite{Sakalli:2016mnk}; the difference can be traced to the two-horizon structure of the Kerr-EMDA geometry, where the ratio $r_+/\delta_r > 1$ generically. In the Schwarzschild limit, our result reduces to $\delta S_{\text{BH}} = 4\pi$, corresponding to the area quantum $\delta\mathcal{A}_H = 16\pi\ell_P^2$, in agreement with the Bekenstein-Maggiore result~\cite{Maggiore:2007nq}.

{\color{black}A central observation of the present work is that the HeunC~$\to$~${}_2F_1$ reduction requires only $\mu_s = 0$ \cite{Sakalli:2016mnk} and does not depend on the scalar charge $q$, since the parameters $\tilde{\alpha}$ and $\tilde{\delta}$ are both $q$-independent. This allowed us to derive the first closed-form GF not only for massless neutral scalars~\eqref{GF2} but also for massless \emph{charged} scalars~\eqref{GFcharged} on the Kerr-EMDA BH, with the $q = 0$ result recovered as a special case.} The dilaton parameter enhances the GF at low energies compared to the Kerr case, implying that Kerr-EMDA BHs are more transparent to scalar radiation. This enhancement is consistent with the effective potential analysis, which shows that the dilaton lowers and broadens the scattering barrier while shifting the photon sphere inward (Fig.~\ref{fig:Veff}). The numerical evaluation of the GF confirms a clear partial wave hierarchy $\Gamma_0 \gg \Gamma_1 \gg \Gamma_2 \gg \Gamma_3$ at low energies (Fig.~\ref{fig:GF_ell}), with the dilaton enhancement being most pronounced for the $s$-wave (Fig.~\ref{fig:GF_dilaton}) and higher spin further increasing the transmission (Fig.~\ref{fig:GF_spin}). {\color{black}The charged GF incorporates the electromagnetic coupling through the $q$-dependent Frobenius exponents $\tilde{\beta}$ and $\tilde{\gamma}$, and when combined with the flux-normalized absorption probability~\eqref{GFphysical}, it captures superradiant amplification: for $\omega < \omega_c = m\Omega_H - q\Phi_H$, the absorption probability turns negative, confirming energy extraction from the BH (Table~\ref{tab:superradiance}). Same-sign scalar and BH charges narrow the superradiant window, while for $q > q_{\mathrm{cr}} = m\Omega_H/\Phi_H$ superradiance is quenched entirely.} The GF-weighted Hawking spectrum reveals that Kerr-EMDA BHs radiate at a higher peak frequency and with greater total luminosity than Kerr BHs of identical mass and spin --- a direct consequence of the elevated Hawking temperature and the enhanced transmission (Fig.~\ref{fig:Hawking}). Meanwhile, the absorption cross-section satisfies the expected low-energy universality $\sigma_{\text{abs}} \to \mathcal{A}_H$ and develops dilaton-dependent oscillatory features at intermediate energies that could serve as observational signatures (Fig.~\ref{fig:sigma_abs}).

Several directions for future work suggest themselves.  {\color{black}Extending the GF calculation to massive scalars ($\mu_s \neq 0$) remains an open problem: the irregular singularity at spatial infinity is genuine for $\mu_s \neq 0$, and the transmission coefficient depends on the CHF connection coefficients, which are expressible only as infinite continued fractions. Their numerical evaluation via the Painlev\'{e}/isomonodromic deformation framework~\cite{Bonelli:2021uvf} would complete the analytical programme initiated here.} The spin-1 and spin-2 perturbation equations on the Kerr-EMDA background have not yet been studied analytically; the Teukolsky formalism may admit a CHF separation similar to the scalar case. {\color{black}Connecting the analytical resonant spectrum obtained
here to the numerical QNM calculations for rotating dilatonic
BHs~\cite{Kokkotas:2015uma} and to the full domain of existence
mapped in Ref.~\cite{Herdeiro:2025blx} would provide a bridge
between exact Heun-function methods and the broader landscape
of Einstein-Maxwell-dilaton solutions at general coupling
$\gamma$.} The parameter-dependent entropy quantum~\eqref{deltaSBH2} calls for a microscopic explanation within string theory, where the EMDA BH arises as a heterotic string solution~\cite{Sen:1992ua}. Finally, the enhanced Hawking emission of Kerr-EMDA BHs could have implications for primordial BH evaporation constraints in scenarios where the dilaton field is light.

} 

\section*{Acknowledgments} 
{\color{black}We thank R.~A.~Konoplya, A.~Zhidenko, and
E.~dos~Santos~Costa~Filho for drawing our attention to related
work on dilaton BH perturbations and extremal solutions.} {\color{black}We are grateful to the Handling Editor, Prof.~Dr.~Jutta Kunz,
and to the anonymous Referee for constructive comments and suggestions
that improved the presentation of this work.}
\.{I}.~S. is thankful for academic support provided by EMU,
T\"{U}B\.{I}TAK, ANKOS, and SCOAP3, as well as for networking support
received through COST Actions CA22113, CA21106, CA23130, and CA23115.
{\color{black}The supplementary Maple file verifying the analytical results
is available alongside the published article.}

\section*{Conflict of Interest}
Authors declare(s) no such conflict of interest.

\appendix

\section{Derivation of Heun Parameters for the Charged Massive KGE} \label{appA}

In this appendix, we provide the detailed derivation of the five CHE parameters listed in Sec.~\ref{sec3e}. We start from the radial ODE~\eqref{radialODE} in the coordinate $\zeta = (r - r_-)/\delta_r$ introduced in Eq.~\eqref{zetadef}. In terms of $\zeta$, $\Delta = \delta_r^2\,\zeta(\zeta-1)$, $r = r_- + \delta_r\,\zeta$, and
\begin{equation}\label{app_Delta}
\frac{d}{dr} = \frac{1}{\delta_r}\frac{d}{d\zeta}\,, \qquad \Delta\frac{d}{dr} = \delta_r\,\zeta(\zeta-1)\frac{d}{d\zeta}\,.
\end{equation}
The radial equation~\eqref{radialODE} becomes
\begin{equation}\label{app_radial}
\delta_r^2\,\frac{d}{d\zeta}\!\left[\zeta(\zeta-1)\frac{dR}{d\zeta}\right] + \left[\frac{\mathcal{K}^2(\zeta)}{\delta_r^2\,\zeta(\zeta-1)} + 2am\omega - \lambda_{\ell m} - \mu_s^2\,\Sigma_r(\zeta) - a^2\omega^2\right]R = 0\,.
\end{equation}
Dividing by $\delta_r^2$ and expanding the derivative:
\begin{equation}\label{app_radial2}
\zeta(\zeta-1)R'' + (2\zeta-1)R' + \left[\frac{\mathcal{K}^2(\zeta)}{\delta_r^4\,\zeta(\zeta-1)} + \frac{2am\omega - \lambda_{\ell m} - \mu_s^2\,\Sigma_r(\zeta) - a^2\omega^2}{\delta_r^2}\right]R = 0\,.
\end{equation}

The function $\mathcal{K}(\zeta) = \mathcal{K}_- + (2M\omega+qQ)\delta_r\,\zeta$ is linear in $\zeta$, so $\mathcal{K}^2/[\zeta(\zeta-1)]$ has simple poles at $\zeta = 0$ and $\zeta = 1$. We decompose using partial fractions and identify the Frobenius indices at $\zeta = 0$ and $\zeta = 1$ from the coefficients of the $1/\zeta$ and $1/(\zeta-1)$ poles.

Substituting the ansatz~\eqref{Rsubstitution}, $R = \zeta^{\tilde{\beta}/2}(\zeta-1)^{\tilde{\gamma}/2}e^{\tilde{\alpha}\zeta/2}F$, and demanding that the singularities at $\zeta = 0$, $\zeta = 1$, and $\zeta \to \infty$ are absorbed by the prefactors, we obtain:

\medskip
\noindent\emph{Inner horizon} ($\zeta = 0$): The indicial equation gives $s_-(s_--1) + s_- + \mathcal{K}_-^2/\delta_r^2 = 0$, so $s_- = \pm i\mathcal{K}_-/\delta_r$. Choosing the sign for ingoing waves: $\tilde{\beta} = 2s_- = 2i\mathcal{K}_-/\delta_r$, yielding Eq.~\eqref{betatilde}.

\medskip
\noindent\emph{Outer horizon} ($\zeta = 1$): Similarly, $\tilde{\gamma} = 2s_+ = -2i\mathcal{K}_+/\delta_r$, yielding Eq.~\eqref{gammatilde}. The negative sign is required for the ingoing boundary condition at the event horizon.

\medskip
\noindent\emph{Spatial infinity} ($\zeta\to\infty$): The leading behavior is $R\sim e^{\pm\sqrt{\mu_s^2-\omega^2}\delta_r\,\zeta}$. The decaying solution requires $\tilde{\alpha}/2 = -\sqrt{\mu_s^2-\omega^2}\,\delta_r$, giving Eq.~\eqref{alphatilde} (with $\sqrt{\mu_s^2-\omega^2} = i\sqrt{\omega^2-\mu_s^2}$ for $\omega > \mu_s$).

\medskip
The parameters $\tilde{\delta}$ and $\tilde{\eta}$ are then obtained by matching the remaining polynomial terms in $\zeta$ after the substitution. Writing $F = \text{HeunC}(\tilde{\alpha},\tilde{\beta},\tilde{\gamma},\tilde{\delta},\tilde{\eta};\zeta)$ and comparing with the canonical CHE~\eqref{CHE}, we read off $\tilde{\delta}$ from the coefficient of $\zeta$ in the numerator of the $F'/[\zeta(\zeta-1)]$ equation, and $\tilde{\eta}$ from the constant term. The resulting expressions are Eqs.~\eqref{deltatilde} and~\eqref{etatilde}.

As a cross-check, the consistency relation~\eqref{paramcheck} follows directly from $\tilde{\beta}+\tilde{\gamma} = 2i(\mathcal{K}_- - \mathcal{K}_+)/\delta_r = -2i(2M\omega+qQ)$, where we used $\mathcal{K}_+ - \mathcal{K}_- = (2M\omega+qQ)\delta_r$.

\section{Confluent Heun Function: Polynomial Condition and ${}_2F_1$ Reduction} \label{appB}

\subsection*{B.1 \; Confluent Heun function}

The confluent Heun function $\text{HeunC}(\alpha,\beta,\gamma,\delta,\eta;\zeta)$ is defined as the solution of the CHE~\eqref{CHE} that is regular at $\zeta = 0$ and normalized such that $\text{HeunC}(\alpha,\beta,\gamma,\delta,\eta;0) = 1$~\cite{Ronveaux:1995}. Its power series representation is
\begin{equation}\label{HeunC_series}
\text{HeunC}(\alpha,\beta,\gamma,\delta,\eta;\zeta) = \sum_{k=0}^{\infty} c_k\,\zeta^k\,, \qquad c_0 = 1\,,
\end{equation}
where the coefficients satisfy the three-term recurrence relation
\begin{equation}\label{recurrence}
A_k\,c_{k+1} + B_k\,c_k + C_k\,c_{k-1} = 0\,, \qquad k = 0,1,2,\ldots
\end{equation}
with $c_{-1} = 0$ and
\begin{align}
A_k &= (k+1)(k+1+\beta)\,,\\
B_k &= -\left[k(k+\alpha+\beta+\gamma+1) + \tfrac{1}{2}\alpha(\beta+1) + \eta\right]\,,\\
C_k &= \tfrac{1}{2}\alpha(k-1+\beta+\gamma) + \delta + \tfrac{1}{2}\alpha\,.
\end{align}

\subsection*{B.2 \; Polynomial condition ($\varepsilon_n$)}

The series~\eqref{HeunC_series} truncates to a polynomial of degree $n$ if and only if two conditions are satisfied simultaneously:
\begin{equation}\label{truncation}
c_{n+1} = 0 \qquad\text{and}\qquad C_{n+1} = \tfrac{1}{2}\alpha(n+\beta+\gamma) + \delta + \tfrac{1}{2}\alpha = 0\,.
\end{equation}
The second condition (on $C_{n+1}$) is the $\varepsilon_n$-condition used in Sec.~\ref{sec4a}. Rearranging:
\begin{equation}\label{epscondition_app}
\frac{\delta}{\alpha} + \frac{\beta+\gamma}{2} + 1 = -n\,,
\end{equation}
which is Eq.~\eqref{epsiloncondition}. The first condition ($c_{n+1}=0$) imposes an additional algebraic constraint that selects the allowed values of the accessory parameter $\eta$ (or equivalently $\lambda_{\ell m}$); for the asymptotic analysis of Sec.~\ref{sec4}, only the $\varepsilon_n$-condition is needed.

\subsection*{B.3 \; Reduction to ${}_2F_1$}

When the irregular singularity at $\zeta = \infty$ degenerates to a regular singularity, the CHE reduces to a Fuchsian equation with three regular singular points, which is the Gauss hypergeometric equation. {\color{black}In our problem, when $\mu_s = 0$ (for arbitrary $q$,
since $\tilde{\delta}$ is $q$-independent; see Sec.~\ref{sec6e}),
the parameter $\tilde{\delta}$ becomes proportional to $\tilde{\alpha}$:}
\begin{equation}\label{delta_alpha_relation}
\tilde{\delta}\big|_{\mu_s=0} = \tilde{\alpha}\!\left[-1-(M+D) + \frac{2am\omega - \lambda_{\ell m} - a^2\omega^2}{2\omega^2\delta_r^2}\right]\,,
\end{equation}
so that $\tilde{\delta}/\tilde{\alpha}$ is a finite constant. Changing to the variable $z = 1-\zeta$ and absorbing the $e^{\tilde{\alpha}\zeta/2}$ factor through a gauge transformation, the CHE reduces to
\begin{equation}\label{reduced_2F1}
z(1-z)F'' + [c_0 - (a_0+b_0+1)z]F' - a_0\,b_0\,F = 0\,,
\end{equation}
with the hypergeometric parameters $a_0$, $b_0$, $c_0$ given in Eqs.~\eqref{a0}--\eqref{c0}. The solution regular at $z = 0$ (the outer horizon) is ${}_2F_1(a_0,b_0;c_0;z)$, from which the GF is obtained by analytic continuation to $z\to -\infty$ as detailed in Sec.~\ref{sec6b}.

\bibliography{ref}
\bibliographystyle{apsrev4-1}

\end{document}